%% file: ms.tex
\pdfoutput=1

\documentclass[journal]{IEEEtran}
\usepackage{cite}
\usepackage[hidelinks]{hyperref}

\ifCLASSINFOpdf
\else
\fi
\input{packages.tex}

\input{colordef.tex}                          
\input{commands.tex}

\input{theorems.tex}
\input{tikzdef.tex}   
\makeatletter
\newcommand\notsotiny{\@setfontsize\notsotiny\@vipt\@viipt}
\makeatother

\newcommand{\citep}[1]{\cite{#1}}
\begin{document}
\title{Stabilizing Quasi-Time-Optimal Nonlinear Model Predictive Control with Variable Discretization}

\author{Christoph~R\"osmann, %,~\IEEEmembership{Member,~IEEE,}
        Artemi~Makarow, %~\IEEEmembership{Fellow,~OSA,}
        and~Torsten~Bertram %,~\IEEEmembership{Life~Fellow,~IEEE}% <-this % stops a space
\thanks{The authors are with the Institute of Control Theory and Systems Engineering, TU Dortmund University, D-44227, Germany, (e-mail: christoph.roesmann@tu-dortmund.de).}% <-this % stops a space
%\thanks{Manuscript received April 19, XXXX; revised August 26, XXXX.}
}

\maketitle

\begin{abstract}
This paper deals with the development and analysis of novel time-optimal point-to-point model predictive control concepts for nonlinear systems.
%Common realizations of model predictive controllers apply direct transcription methods to first discretize and then optimize the subordinate optimal control problems.
Recent approaches in the literature apply a time transformation, however, which do not maintain recursive feasibility for piecewise constant control parameterization.
The key idea in this paper is to introduce uniform grids with variable discretization. % in combination with a shrinking horizon grid adaption scheme.
%The minimization of the local time information of a grid leads to an overall time-optimal transition.
A shrinking-horizon grid adaptation scheme ensures convergence to a specific region around the target state and recursive feasibility. 
The size of the region is configurable by design parameters. This facilitates the systematic dual-mode design for quasi-time-optimal control to restore asymptotic stability and establish a smooth stabilization.
Two nonlinear program formulations with different sparsity patterns are introduced to realize and implement the underlying optimal control problem.
For a class of numerical integration schemes, even nominal asymptotic stability and true time-optimality are achieved without dual-mode.
A comparative analysis as well as experimental results demonstrate the effectiveness of the proposed techniques. % TODO: falls die experimental results dirn bleiben
\end{abstract}

\begin{IEEEkeywords}
Predictive control, Minimum-time control, Time-optimal control, Direct transcription, Variable discretization, Hypergraph, Dual-mode
\end{IEEEkeywords}

% \fi
\IEEEpeerreviewmaketitle

%\section{Introduction}
%\IEEEPARstart{T}{his} demo file is intended to serve as a ``starter file''
%for IEEE journal papers produced under \LaTeX\ using
%IEEEtran.cls version 1.8b and later.
%I wish you the best of success.

\input{introduction.tex}

\input{theory.tex}

\input{example.tex}

\input{summary.tex}

% or


%\appendices

%\input{appendix.tex}

%% you can choose not to have a title for an appendix
%% if you want by leaving the argument blank
%\section{}

\section*{Acknowledgment}

This work is funded by the German Research Foundation (DFG, \mbox{BE 1569/13-1}).  % here.

\ifCLASSOPTIONcaptionsoff
  \newpage
\fi

\bibliography{literature}

\end{document}

%% file: packages.tex
\usepackage[super]{nth} % \nth{1}, \nth{2}, \nth{3}, \nth{4}

\usepackage{amsmath}%
\usepackage{amsfonts}%
\usepackage{amssymb}%
\usepackage{mathtools}%
\usepackage{mathrsfs} % usage: $\mathscr{ABCDEFGHIJKLMNOPQRSTUVWXYZ}$

\usepackage{tabularx}
\usepackage{booktabs}
\usepackage[table]{xcolor}
\newcolumntype{Y}{>{\centering\arraybackslash}X}
\usepackage[exponent-product = \cdot,
			output-complex-root = j, 
			separate-uncertainty = true,
			output-product = \cdot,
			arc-separator = \,,
			product-units = brackets-power]{siunitx}
\usepackage{subcaption}

\usepackage{tikz,pgfplots}

\usepackage{varwidth} % helpful for comments in algorithms

%% file: colordef.tex
\definecolor{black}{rgb}{0.0,0.0,0.0}
\definecolor{darkgray}{rgb}{0.48,0.48,0.48}
\definecolor{darkgraytext}{rgb}{0.32,0.32,0.32}
\definecolor{lightgray}{rgb}{0.70,0.70,0.70}

\definecolor{BlueColor}{rgb}{0.00000,0.44700,0.74100} % Helligkeit mit 0.396 zu gering

%% file: commands.tex
\def\MPC{MPC}

\def\OCP{optimal control problem}

\newcommand*{\mtext}[1]{\text{\normalfont #1}} % enforce non-italic also in theorem environments

\newcommand*{\x}{\ensuremath{x}}%
\newcommand*{\xc}[1]{\ensuremath{\x(#1)}}%
\newcommand*{\xcdot}[1]{\ensuremath{\dot{\x}(#1)}}%
\newcommand*{\xref}{\ensuremath{\x_\text{ref}}}%

\newcommand*{\xtilde}{\ensuremath{\tilde{\x}}}%

\newcommand*{\xlin}{\ensuremath{\x_\mtext{lin}}}%
\newcommand*{\xlindot}{\ensuremath{\dot{\x}_\mtext{lin}}}%
\newcommand*{\ulin}{\ensuremath{\u_\mtext{lin}}}%

\newcommand*{\uref}{\ensuremath{\u_\mtext{f}}}%
\newcommand*{\urefscalar}{\ensuremath{u_\mtext{f}}}%

\renewcommand*{\u}{\ensuremath{u}}%
\newcommand*{\uc}[1]{\ensuremath{\u(#1)}}%

\newcommand*{\xs}{\ensuremath{\x_{\mtext{s}}}}%
\newcommand*{\ts}{\ensuremath{t_{\mtext{s}}}}%
\newcommand*{\xsolfun}{\varphi}
\newcommand*{\xmusolfun}{\varphi_\mu}
\newcommand*{\f}{\ensuremath{f}}

\newcommand*{\xcel}[2]{\ensuremath{\x_{#2} (#1)}}% 
\newcommand*{\xud}[1]{\ensuremath{\x_{#1}}}% 
\newcommand*{\uuc}[1]{\ensuremath{\u(#1)}}%
\newcommand*{\uud}[1]{\ensuremath{\u_{#1}}}%

\newcommand*{\uustar}{\ensuremath{\u^*}}%
\newcommand*{\uucstar}[1]{\ensuremath{\u^*(#1)}}%
\newcommand*{\uucstarbig}[1]{\ensuremath{\u^*\big(#1\big)}}%
\newcommand*{\uudstar}[1]{\ensuremath{\u^*_{#1}}}%

\newcommand*{\vardt}{\ensuremath{\Delta t}}
\newcommand*{\vardtstar}{\ensuremath{\Delta t^*}}
\newcommand*{\vardtref}{\ensuremath{\vardt_\mtext{s}}}
\newcommand*{\vardthyst}{\ensuremath{\vardt_\epsilon}}
\newcommand*{\vardtmax}{\ensuremath{\vardt_\mtext{max}}}

\newcommand*{\vardtmin}{\ensuremath{\vardt_\mtext{min}}}

\newcommand*{\Nmax}{\ensuremath{N_\mtext{max}}}%
\newcommand*{\Nmin}{\ensuremath{N_\mtext{min}}}%
\newcommand*{\xmu}{\ensuremath{\x_\mu}}%
\newcommand*{\xmuc}[1]{\ensuremath{\xmu{(#1)}}}%
\newcommand*{\xmucdot}[1]{\ensuremath{\dot{\x}_\mu(#1)}}%

\newcommand*{\xmucel}[2]{\ensuremath{\x_{\mu,#2} (#1)}}% 

\newcommand*{\ecpsf}{\ensuremath{\xfel{1}}}% ECP POSITION

\newcommand*{\xf}{\ensuremath{\x_\mtext{f}}}%
\newcommand*{\xfel}[1]{\ensuremath{\x_{\mtext{f},#1}}}%

\newcommand*{\xeq}{\ensuremath{\xf}}%

\newcommand*{\tmu}{\ensuremath{t_\mu}}%
\newcommand*{\tmud}[1]{\ensuremath{t_{\mu,#1}}}%
\newcommand*{\tf}{\ensuremath{t_\mtext{f}}}
\newcommand*{\tfstar}{\ensuremath{t^*_\mtext{f}}}
\newcommand*{\tfmin}{\ensuremath{t_\mtext{min}}}

\newcommand*{\xmus}{\ensuremath{\x_{\mtext{s}}}}%
\newcommand*{\tmus}{\ensuremath{\tmud{0}}}%

\newcommand*{\mulaw}{\ensuremath{\mu}}%
\newcommand*{\mulawlin}{\ensuremath{\mu_\mtext{lin}}}%
\newcommand*{\mulawdual}{\ensuremath{\mu_\mtext{dual}}}%

\newcommand*{\vardtmud}[1]{\ensuremath{\Delta t_{\mu,#1}}}

\newcommand*{\nullvec}{\ensuremath{0}}%

\newcommand*{\realpos}{\ensuremath{\mathbb{R}^+}}%
\newcommand*{\realposzero}{\ensuremath{\mathbb{R}^+_0}}%
\newcommand*{\naturalzero}{\ensuremath{\mathbb{N}_0}}%
\newcommand*{\naturalpos}{\ensuremath{\mathbb{N}}}% without zero
\newcommand*{\xset}{\ensuremath{\mathcal{X}}}%
\newcommand*{\rxset}{\ensuremath{\mathbb{X}}}%
\newcommand*{\uset}{\ensuremath{\mathcal{U}}}%
\newcommand*{\ruset}{\ensuremath{\mathbb{U}}}%
\newcommand*{\upcset}[1]{\ensuremath{\mathcal{U}^N(#1)}}%
\newcommand*{\uadmset}[2]{\ensuremath{\mathscr{U}^N(#1,#2)}}%
\newcommand*{\Pfull}[2]{\ensuremath{P_{c}(#1, #2)}}%
\newcommand*{\Pfullbig}[2]{\ensuremath{P_{c}\big(#1, #2\big)}}%
\newcommand*{\Pxfshrinking}{\ensuremath{P_{c}\big(\Nmin, (\Nmin-1)\vardtstar(\xs,N)\big)}}%

\newcommand*{\xfset}{\ensuremath{\mathbb{X}_\mtext{f}}}%

\newcommand*{\rxsetlin}{\ensuremath{\mathbb{X}_\mtext{lin}}}%

\newcommand*{\optimparamglobal}[1]{\ensuremath{\mathcal{P}_\mtext{global}}}

\newcommand*{\tcpu}{\ensuremath{\Delta t_\mtext{cpu}}}
\newcommand*{\tcpumed}{\ensuremath{\Delta t_\mtext{cpu}}}

\newcommand*{\ierrtf}{\ensuremath{e_{\hat{x}}}}

\DeclareMathOperator{\nablaop}{\nabla}
\newcommand*{\gradwrt}[2]{\ensuremath{\nablaop_{\kern -0.2em #2} #1}} % adjust horizontal spacing https://tex.stackexchange.com/questions/432200/ugly-horizontal-spacing-with-some-symbol-subscript-superscript-combinations
\DeclareMathOperator{\Djac}{D}

\newcommand*{\jacobwrt}[2]{\ensuremath{\Djac_{#2} #1}}

\newcommand*{\hessianwrt}[2]{\ensuremath{\nablaop^2_{\kern -0.2em #2} #1}}

\newcommand*{\KLfun}{\ensuremath{\mathscr{K\kern -0.3em L}}}

\newcommand{\dtau}{\text{d}\tau}

\newcommand{\transpose}{\intercal}

\newcommand{\fundef}[3]{#1\,{:}\,#2\,{\mapsto}\,#3}
\newcommand{\mdef}{\vcentcolon=}

\newcommand{\setmuskip}[2]{#1=#2\relax} % https://tex.stackexchange.com/questions/188215/mathtools-with-mu-illegal-unit/188221#188221

\DeclarePairedDelimiter\abs{\lvert}{\rvert}%
\DeclarePairedDelimiter\norm{\lVert}{\rVert}%

\makeatletter
\let\oldabs\abs
\def\abs{\@ifstar{\oldabs}{\oldabs*}}
\let\oldnorm\norm
\def\norm{\@ifstar{\oldnorm}{\oldnorm*}}
\makeatother

\newcommand{\reffig}[1]{Figure~\ref{#1}}
\newcommand{\reftab}[1]{Table~\ref{#1}}

\newcommand{\refsec}[1]{Section~\ref{#1}}

\newcommand{\refprop}[1]{Proposition~\ref{#1}}
\newcommand{\refrem}[1]{Remark~\ref{#1}}

\newcommand{\refdef}[1]{Definition~\ref{#1}}
\newcommand{\reflem}[1]{Lemma~\ref{#1}}

\newcommand{\reffigc}[1]{Figure~\ref{#1}}

\newcommand{\refsecc}[1]{Section~\ref{#1}}

%% file: theorems.tex
\newtheorem{thm}{Theorem}
\newtheorem{prop}{Proposition}
\newtheorem{cor}{Corollary}
\newtheorem{lem}[thm]{Lemma}
\newtheorem{rem}{Remark}
\newtheorem{defn}{Definition}

%% file: tikzdef.tex
\usepgfplotslibrary{patchplots}
\usetikzlibrary{positioning,automata,through,math,calc,plotmarks,shapes,arrows,arrows.meta,quotes,decorations.markings,shapes.misc,shapes,backgrounds,patterns,angles,babel,intersections,decorations,spy,decorations.pathreplacing, shadows,shadings} % include some useful tikz
\pgfplotsset{compat=newest}
\pgfplotsset{plot coordinates/math parser=false}

\pgfplotsset{/pgf/number format/.cd,1000 sep={}} % Trennzeichen bei 1000

\newlength\figureheight
\newlength\figurewidth

\tikzstyle{every picture}+=[remember picture]
\tikzstyle{arrow} = [->,>=stealth']
\tikzstyle{arrowreverse} = [<-,>=stealth']
\tikzstyle{arrowbidir} = [<->,>=stealth']

\tikzset{draw/.append style={line width= 0.3mm}}

\tikzset{thin/.style ={line width= 0.2mm}}
\tikzset{thick/.style ={line width= 0.4mm}}
\tikzset{very thick/.style ={line width= 0.5mm}}

\pgfplotsset{tick style={black}} % modifies the style `every tick'     $ very thin,

\pgfplotsset{every axis plot post/.append style={line join=round}}

\tikzset{every picture/.style={font issue=\small},
	font issue/.style={execute at begin picture={#1\selectfont}}
}
\tikzset{fontscale/.style = {font=\small}
}

\pgfplotsset{every axis/.append style={
		axis lines = left % middle
	}
}

\newlength\BlockSep
\newlength\BlockHeight
\newlength\BlockWidth
\newlength\sepsepsep
\tikzset{%
	partially dashed/.style={
		decoration={show path construction, 
			lineto code={
				\draw[] (\tikzinputsegmentfirst) --($(\tikzinputsegmentfirst)!#1!(\tikzinputsegmentlast)$);,
				\draw[dashed, dash phase=3pt,-latex] ($(\tikzinputsegmentfirst)!#1!(\tikzinputsegmentlast)$)--(\tikzinputsegmentlast);,
			}
		},
		decorate
	},
}

\tikzstyle{vecArrow} = [thick, decoration={markings,mark=at position 1 with {\arrow[semithick]{open triangle 60}}},
double distance=1.4pt, shorten >= 5.5pt, preaction = {decorate}, postaction = {draw,line width=1.4pt, white,shorten >= 4.5pt}]
\tikzstyle{vecInnerWhite} = [semithick, white,line width=1.4pt, shorten >= 4.5pt]

\tikzstyle{vecArrowNL} = [thick, shorten <= 0.75pt, decoration={markings,mark=at position 1 with {\arrow[semithick]{open triangle 60}}},
double distance=1.4pt, shorten >= 5.5pt, preaction = {decorate}, postaction = {draw,line width=1.4pt, white,shorten >= 4.5pt, shorten <= 0.75pt}]
\tikzstyle{vecInnerWhiteNL} = [semithick, white,line width=1.4pt, shorten >= 4.5pt, shorten <= 0.75pt]

\tikzset{%
	block/.style    = {draw, thick, rectangle, minimum height = \BlockHeight, minimum width = \BlockWidth,inner sep=0},
	square/.style    = {draw, thick, rectangle, minimum height = \BlockHeight, minimum width = \BlockHeight},
	nonlinear/.style= {draw, thick, double, rectangle, minimum height = \BlockHeight, minimum width = \BlockWidth},
	sum/.style      = {draw, circle, node distance = 2cm}, % Adder
	branch/.style      = {draw, circle, fill=black, minimum height = 0.4em, inner sep=0pt,anchor=center}, % Adder
	input/.style    = {coordinate}, % Input
	output/.style   = {coordinate}, % Output
	annotation/.style={coordinate}, % Output
	skip loop/.style={to path={-- ++(0,-1.2\BlockSep) -| (\tikztotarget)}},
	skip loop up/.style={to path={-- ++(0, 1.2\BlockSep) -| (\tikztotarget)}},
	point/.style    = {coordinate, anchor=center}
}

\def\centerarc[#1](#2)(#3:#4:#5)% Syntax: [draw options] (center) (initial angle:final angle:radius)
{ \draw[#1] ($(#2)+({#5*cos(#3)},{#5*sin(#3)})$) arc (#3:#4:#5); }

\makeatletter
\pgfplotsset{ % Korrektur der Legenden, die unter dem Plot angezeigt werden (legend pos = outer south /outer north)
	every axis x label/.append style={
		alias=current axis xlabel,
	},
	legend pos/outer south/.style={
		/pgfplots/legend style={
			at={%
				(%
				\@ifundefined{pgf@sh@ns@current axis xlabel}%
				{xticklabel cs:0.5}%
				{current axis xlabel.south}%
				)%
			},
			anchor=north,
			legend columns=3,
			font=\small,
			fill=none,
			/tikz/every even column/.append style={column sep=10pt, font=\small} %Increase spacing in legend
		}
	},
	legend pos/outer north/.style={
		/pgfplots/legend style={
			at={(\figurewidth/2,\figureheight+0.2cm)},
			draw=none, % ignore box around legend
			anchor=south,
			legend columns=3,
			font=\scriptsize,
			fill=none,
			/tikz/every even column/.append style={column sep=10pt, font=\scriptsize} %Increase spacing in legend
		}
	},
	legend pos/outer north2/.style={
		/pgfplots/legend style={
			at={(\figurewidth/2,\figureheight+0.5cm)},
			draw=none, % ignore box around legend
			anchor=south,
			legend columns=3,
			font=\scriptsize,
			fill=none,
			/tikz/every even column/.append style={column sep=5pt, font=\scriptsize} %Increase spacing in legend
		}
	}
}
\makeatother

\newlength\customshift 

%% file: introduction.tex
% !TeX spellcheck = en_US
\section{Introduction}
\label{sec:introduction}

Minimizing time plays a vital role in increasing the productivity of automation solutions  in a variety of industries. % TODO examples
To give a few examples, in the field of warehouse robotics, mobile robots are expected to navigate as fast as possible while avoiding obstacles.
The productivity in the area of automated assembly at automobile manufacturers correlates strongly with the execution speed of their robotic manipulators.
Also, racing is dedicated to minimizing lap times as a central objective.
%Time-optimal feedback control is often achieved by aggressive tuning of conventional controllers. % ref

A comprehensive and generic framework to explicitly account for performance criteria during feedback control is nonlinear model predictive control (MPC) \citep{gruene2017_book, rawlings2017_book}.
%Nonlinear model predictive control (MPC) constitutes an established and comprehensive framework to regulate nonlinear dynamic systems under performance criteria and the presence of constraints 
% TODO:
%In each sampling interval, the controller employs a dynamic model of the plant to predict and optimize the future evolution within a specified temporal horizon.
%The first portion of the optimized control sequence is commanded to the plant before the optimization process is repeated.
Researches in the context of \MPC{} mainly consider quadratic cost terms as performance criteria, in particular, the minimization of the state error and control effort. % TODO Literatur + stability
In the recent years, also the study of economic \MPC{} schemes with arbitrary performance criteria has received major attention. % TODO Literatur + stability
However, the theoretical foundation and findings do not necessarily include minimum-time formulations as these usually require non-fixed final times in the prediction horizon. %considering the variable final time of the prediction horizon is challenging.
The literature mentions dedicated time-optimal \MPC{} realizations rarely.
% TODO: Schwung zur Optimalisteuerung mit zeitoptimalität
Nevertheless, some contributions and applications rely on a time transformation from feedforward optimal control~\cite{quintana1973_joc,teo1999_joams}. % \cite{quintana1973_joc, jennings1991_aiesw, teo1991_book}. % TODO was ist mit quadratic form und aggressivem tuning?
Hereby, the variable grid is mapped on to a fixed grid in a new time scale and accordingly the system dynamics equation is scaled by additional optimization parameters.
%A direct method for bang-bang control and first- and second-order conditions as well as first- and second-order variational derivatives
%of the state trajectory with respect to the switching times are provided in~\cite{vossen2010_jota}. 
% TODO vorheriges Zitat bringne?
Stability results for controllers considering the time transformation with simple state feedback are still intractable, especially due to the control parameterization applied in direct optimal control and~\MPC{}, and are hence not yet available in the literature (see also~\cite{verschueren2017_cdc}). %  -- to the author's best knowledge --
Zhao et al.~provide a time-optimal \MPC{} scheme for the control of a spherical robot based on the time transformation~\cite{zhao2004_asce}.
A hybrid cost function that also considers quadratic form cost achieves stabilization.
The transformed time is bounded from below close to the target state such that only the quadratic form cost becomes active. 
However, the approach does not guarantee recursive feasibility.
Verschueren et al. compute time-optimal motions along a Cartesian path for robotic manipulators~\cite{verschueren2016_acc}.
Time transformation is applied to the underlying time-optimal control problem to map states and controls onto a fixed integration grid.
In applications such as race car automatic control, tailored \MPC{} methods minimize the lap time \cite{kelly2010_vsd,timings2010_isavc,verschueren2014_cdc,verschueren2016_ecc}. % TODO prüfen was jetzt MPC ist und was nur open-loop
%Explicit \MPC{} is extended to the offline computation of time-optimal tasks for linear time-invariant and piecewise affine systems \cite{grieder2005_automatica}.
%An approximation based on Voronoi diagrams for nonlinear systems is provided in~\cite{raimondo2012_ifac}.

% TODO drawbacks time transformation (z.B. konvergenz-/ stabilitätsnachweise, adapation des grids, ...)

A nonlinear \MPC{} method for time-optimal point-to-point transitions which does not rely on time transformation is presented in~\cite{vandenbroeck2011_mechatronics,vandenbroeck2011_ifac}. % TODO hier gibt es noch weitere referenzen
The method called TOMPC minimizes the settling time, i.e. horizon length $N$, in a two-layer optimization routine.
%The settling time is defined in terms of the horizon length $N$.
The outer loop incrementally decreases $N$ until the inner loop nonlinear program with a standard quadratic form cost fails to generate a feasible solution for the allocated time horizon.
Since the cost function minimizes the distance of discrete states to the final state, the solution with the shortest feasible horizon is quasi time-optimal. 
Due to the lower bound on the time horizon, the algorithm behaves like a conventional \MPC{} in the vicinity of the final state and therefore guarantees asymptotic stability. 
The computation time strongly depends on the initial estimate of the settling time, as it determines the number of iterations in the outer loop time horizon reduction.  %As part of the evaluation of the approach presented in this paper, we use TOMPC for a comparison.
%Refer to \refapp{app:related_to_methods:tompc} for a mathematically more detailed description of TOMPC.
Properties on time-optimal MPC for discrete-time systems in general are discussed in~\cite{sutherland2019_cdc}.
An alternative approach that follows a reference path in minimum time %, rather than performing a point-to-point transition,
is presented in \cite{lam2012_phd}. Time-optimality is nearly achieved in case of long time horizons. 
A time-optimal approach for linear systems is presented in~\citep{besselmann2009_cdc}.

An approach that considers $\ell_1$-norm cost functions for linear systems is described in~\cite{homsi2016_phd}.
For general nonlinear systems, Verschueren et al. proposes a stabilizing time-optimal \MPC{} approach based on a weighted $\ell_1$-norm cost~\cite{verschueren2017_cdc}.
The approach considers discrete-time respectively sampled-data models and guides the system towards a target state in minimum time and stabilizes it there.
It is required that the horizon length $N$ is sufficiently large such that the target state is reachable within $N$ time steps.
The single-stage optimization as well as milder assumptions on $N$ are superior in comparison to TOMPC. % TODO: bessere beschreibung
Since the $\ell_1$-norm is non-smooth, it is replaced by a smooth representation in every practical implementation. 
It consists of additional slack variables which might increase computation times significantly for larger problems.
In addition, a weighting parameter must be chosen properly to ensure time-optimality while maintaining numerical well-conditioning of the underlying optimization problem.
%A more detailed description is provided in \refapp{app:related_to_methods:l1norm}.

% TODO contribution (hier: point-to-point anstatt path following)
Previous work proposes time-optimal point-to-point \MPC{} formulations based on direct transcription and variable discretization~\cite{roesmann2015_ecc, roesmann2017_cdc}.
A dedicated grid adaptation scheme adjusts the temporal resolution with respect to a predefined sample time during runtime. 
However, the presented work does not take stability and recursive feasibility maintenance and guarantees into account.
To this end, these important challenges are addressed in this paper and the novel contributions are as follows:
Time-optimal \MPC{} with variable discretization is formulated based on sampled-data systems with piecewise constant control parameterization, instead of relying on direct transcription initially~\cite{roesmann2019_phd}.
A shrinking horizon grid adaptation scheme enables the derivation of convergence and recursive feasibility results. % based on Lyapunov functions and the dynamic programming principle. % TODO der letzte teil könnte gekürzt werden
The size of the target region which is guaranteed to be reached is adjusted by design of the controller, e.g. by choosing a proper grid size. 
In addition, the derivation of these results include lower bounds on the temporal resolution and grid size, which are often indispensable in practice.
The theoretical results either allow \textit{a true time-optimal point-to-point transition} to adhere to predefined tolerances or enable \textit{a systematic quasi-time-optimal dual-mode controller design} which restores true asymptotic stability.
%To our best knowledge, these results are not yet available in the literature.
%Even with the time transformation, in which the switching of the controller is already applied before reaching the steady state, recursive feasibility is not maintained with piecewise constant controls.
%Instead, the techniques proposed in this paper do.
Finally, two nonlinear program formulations with different sparsity patterns are presented that mimic the optimal control problem.
Depending on the underlying numerical integration scheme and stricter conditions, even asymptotic stability results are derived for the nominal system without dual-mode realization.

The outline of this paper is as follows: \refsec{sec:prelim} introduces some preliminaries and a formal description of the \MPC{} realization.
\refsecc{sec:stability_analysis} discusses stability and recursive feasibility issues, proposes a grid adaptation scheme and presents convergence results.
Details on the practical realization are provided in \refsec{sec:numerical_realization}.
Quasi-time optimal stabilizing control based on dual-mode is described in \refsec{sec:dual-mode}.
\refsecc{sec:example} provides a numerical example and compares the proposed techniques with the state of the art.
A demonstration on a real system is provided in \refsec{sec:ecp}, and finally \refsec{sec:summary} concludes the work. % TODO: falls ECP drin bleibt

% TODO OUTLINE  (vllt. wandert davon noch einiges in die contribution!!)
%The outline of this paper is as follows: \refsec{sec:prelim} introduces some preliminaries and the problem description. 
%This section includes the control parameterization, a formal description of the related minimum-time optimal control problem with variable discretization and its integration with state feedback.
%\refsecc{sec:stability_analysis} discusses stability and recursive feasibility issues occurring from grids with variable discretization. 
%A proposed grid adaptation scheme accounts for those bottlenecks and enables the derivation of practical stability results. 
%In \refsec{sec:dual-mode}, a dual-mode formulation is described which restores asymptotic stability.
%Details on the implementation and realization are provided in \refsec{sec:numerical_realization}, especially sparse nonlinear program formulations of the minimum-time optimal control problem. 
%Under certain conditions, nominal asymptotic stability results for these realizations are presented.
%A numerical example in \refsecc{sec:example} demonstrates the proposed schemes.
%Finally, \refsec{sec:summary} summarizes the results and provides an outlook on future work.

% TODO
%The nonlinear programs in \refchap{chap:uniform_grid} contain the final state as an equality condition similar to the optimal control problems with time transformation for which stability and recursive feasibility are often \textit{assumed} to be valid.

%% file: theory.tex
% !TeX spellcheck = en_US
\section{Preliminaries and Problem Setup}
\label{sec:prelim}

%% Definition of special sets here:
\subsection{Notation}

% TODO: Dieser Abschnitt kann vielleicht weg!

Let $\naturalzero$ denote the set of non-negative integers and $\realposzero$ the set of non-negative real numbers.
Furthermore, $\emptyset$~represents an empty set.
%A function $\fundef{\alpha}{\realposzero}{\realposzero}$ is of class~$\Kfun$ if it is continuous and strictly increasing with $\alpha(0)=0$.
%If in addition $\alpha$ is unbounded, then it is of class $\Kinffun$.
Let $\jacobwrt{\f}{u}(\bar{u})$ denote the Jacobian of $\f$ w.r.t. $\u$ and evaluated at $\bar{u}$.
The set of Lebesgue integrable mappings from interval $I\subseteq \mathbb{R}$ to $\mathbb{R}^q$ is denoted by $L^\infty(I, \mathbb{R}^q)$.

% TODO hinweis: notation angelehnt an grüne?

% TODO: hinweis über closed-loop and open loop notation?

%% Dynamic system:
\subsection{Dynamic System}
\label{sec:dyn_sys}
We consider continuous-time, nonlinear, time-invariant systems with state trajectory $\fundef{\x}{\mathbb{R}}{\xset}$ and control trajectory $\fundef{\u}{\mathbb{R}}{\uset}$:
\begin{equation}
\xcdot{t} = \f\big(\xc{t},\uc{t}\big). %  , \quad \xc{t=0}=\xs
\label{eq:system_dynamics_cont}
\end{equation}
Throughout this paper, the state space $\xset$ and control space $\uset$ are defined as $\xset \mdef \mathbb{R}^p$ and $\uset \mdef \mathbb{R}^q$ with state vector dimension $p\in\naturalpos$ and control vector dimension $q\in\naturalpos$ respectively. 
%Depending on the actually used direct optimal control strategy and optimization algorithm, more general metric spaces can be considered.
Function $\fundef{\f}{\xset \times \uset}{\xset}$ defines a nonlinear mapping of the state and control trajectory, $\xc{t}$ and $\uc{t}$ respectively, to the state velocity $\xcdot{t}$ embedded in $\xset$.
System~\eqref{eq:system_dynamics_cont} is further subject to state and input constraint sets, i.e. $\xc{t} \in \rxset \subseteq \xset$ and $\uc{t} \in \ruset \subset \uset$, respectively.
The solution to~\eqref{eq:system_dynamics_cont} contained in an open time interval $I \subseteq \mathbb{R}$ with initial value $\xc{\ts}=\xs$, $\ts \in I$, $\xs \in \xset$ and $t \in I$ is defined by 
\begin{equation}
\xsolfun\big(t, \xs, \uc{t}\big) = \xs + \int_{\ts=0}^{t} \f\big(\xc{\tau},\uc{\tau}\big) \dtau. % TODO: \xc required?:  \xuc{t} \mdef 
\label{eq:ivp:x}
\end{equation}
Without loss of generality, initial time $\ts$ is fixed to $\ts=0$ as~\eqref{eq:system_dynamics_cont} is time-invariant.
Carath\'{e}odory's existence theorem addresses conditions for the existence and uniqueness of the solution.
In the following, we assume that the vector field $\f$ is continuous and Lipschitz in its first argument.
Furthermore, the control $\uc{t}$ is supposed to be locally Lebesgue integrable for $t \in I$, i.e. $u \in L^\infty(I, \uset)$.%  respectively piecewise continuous 
%Set $L^\infty(I, \uset)$ denotes the corresponding function space.

\subsection{Optimal Control Problem}
In \MPC{}, system~\eqref{eq:system_dynamics_cont} is considered as dynamic model for the underlying \OCP{} to predict the future evolution.
As in the majority of \MPC{} realizations, the control trajectory is parameterized as piecewise constant which also reflects the discrete-time nature of the inherent sampled control law.
To this end, consider the following grid:
$0 = t_0 \leq t_1 \leq \dotsc \leq t_k \leq \dotsc \leq t_N = \tf$ with $t_k,\tf \in I$, $k=0,1,\dotsc,N$ and $N\in\naturalpos$.  % TODO inklusive 0??
Condition $t_{k+1}-t_k = \vardt$ for $k=0,1,\dotsc,N-1$ enforces uniformity with grid partition length $\vardt \in \realposzero$ and hence $t_k = k\vardt$ refers to individual grid points. % TODO: set notation einführen?
Restricting the control trajectory $\uuc{t}$ to constant values $\uud{k}\in\uset$ on each grid partition results in the following control function space:
%\begin{equation}
%\begin{split}
%\mathscr{U}^N \mdef &\{ u \in L^\infty([t_0,\tf], \uset) \mid \exists u_0, u_1,\dotsc,u_{N} \in \uset, \\
%&\exists t_0, t_1,\dotsc,t_{N} \in \uset, \\
%&u(t)\mdef u_k \mtext{ for } t\in[t_k, t_{k+1}), k=0,1,\dotsc,N-1\}.
%\end{split}
%\end{equation}
\begin{equation}
\begin{split}
\upcset{\tf} \mdef &\{ u \in L^\infty([0,\tf], \uset) \mid \exists \, u_0, u_1,\dotsc,u_{N-1} \in \uset, \\
&u(t)\mdef u_k \mtext{ for } t\in[t_k, t_{k+1})  \mtext{ with } t_k = k \tf / N,\\
&k=0,1,\dotsc,N-1\}.
\end{split} % TODO, was ist mit t_N
\end{equation}
%
%TODO hinweis auf point-to-point vs path following oder wurde es schon in der introduction genannt?
The control task considered in this paper comprises the motion from an initial state $\xs$ to a target set $\xfset \subseteq \rxset$ in minimum time. 
In fact, we focus primarily on point-to-point motions such that $\xfset$ is given by $\xfset \mdef \{ \xf \}$ with terminal state $\xf \in \rxset$.
%A control trajectory $u \in \upcset{\tf}$ and the corresponding state trajectory $\xsolfun\big(t,\xs,u(t)\big)$ are called \textit{admissible for $\xs$ up to time $\tf$ with $N$ steps} % TODO steps?
%A control trajectory $u \in \upcset{\tf}$ and the corresponding state trajectory $\xsolfun\big(t,\xs,u(t)\big)$ are called \textit{admissible for $\xs$ up to time $\tf$ with $N$ steps} % TODO steps?
A control trajectory $u \in \upcset{\tf}$ and the corresponding state trajectory $\xsolfun\big(t,\xs,u(t)\big)$ are called \textit{admissible for $\xs$ up to time $\tf$ in $N$ steps} % TODO steps?
if $u(t) \in \ruset$, $\xsolfun\big(t,\xs,u(t)\big) \in \rxset$ for $t\in [0,\tf)$ and $\xsolfun\big(\tf,\xs,u(t)\big) \in \xfset$ hold. 
Accordingly, the function space of all admissible control trajectories is denoted by:
\begin{equation}
\begin{split}
\uadmset{\xs}{\tf} &\mdef \{ u \in \mathcal{U}^N(\tf) \mid u(t) \in \ruset, \xsolfun\big(\tf,\xs,u(t)\big) \in \xfset \\
&\mtext{for } t\in [0,\tf], \xsolfun\big(k \tf /N,\xs,u(t)\big) \in \rxset \mtext{ for}\\
& t\in [0,(k+1) \tf/N) \mtext{ and } k=0,1,\dotsc,N-1\}. % \mtext{ and } 
\end{split}
\label{eq:admissible_set_def}
\end{equation}
%\begin{equation}
%\begin{split}
%\uadmset{\xs}{\tf} \mdef &\{ u \in \mathcal{U}^N(\tf) \mid u(t) \in \ruset, \xsolfun\big(\tf,\xs,u(t)\big) \in \xfset,\\
%&\xsolfun\big(t,\xs,u(t)\big) \in \rxset \mtext{ for } t\in [0,\tf) \}. % \mtext{ and } 
%\end{split}
%\end{equation}
\begin{rem}
	Note, that the admissibility definition~\eqref{eq:admissible_set_def} considers state constraints only at grid points, which is common in sampled-data \MPC{}.
	However, the results in \refsec{sec:stability_analysis} also apply to a definition in which $\xsolfun\big(t,\xs,u(t)\big) \in \rxset$ holds for $t\in [0,\tf)$.
\end{rem}
The \OCP{} of searching for the minimum transition time and corresponding control trajectory is now expressed compactly in mathematical form:
\begin{equation}
	\tfstar(\xs,N) = \underset{u \in \uadmset{\xs}{\tf}, \tf}{\min} \tf \ \ \, \mtext{s.t.} \ N\vardtmin \leq \tf \leq N\vardtmax. % TODO N \dtmin anstatt \tfmin
	\label{eq:ocp}
\end{equation}
Hereby, $\tfstar(\xs,N)$ denotes the minimum transition time and emphasizes its relation to initial state $\xs$ and grid resolution $N$. 
Bounds \mbox{$\vardtmin,\vardtmax \in \realposzero$} with $\vardtmin\leq\vardtmax$ are of a technical nature and their purpose is described later. % TODO: nicht vergessen
We denote the resulting optimal control trajectory by $\uucstar{t, \xs, N}$ and the optimal grid partition length by $\vardtstar(\xs,N) = \tfstar(\xs,N)/N$.
%The additoinal
Note that the grid is time-variable similar to~\cite{roesmann2015_ecc}. % TODO: wirklich ein verweis hier drauf?
% TODO:
The \OCP{}~\eqref{eq:ocp} is referred to as \textit{feasible} if the resulting optimal control and state trajectories are admissible from $\xs$ up to time $\tfstar(\xs,N)$ in $N$ steps.
Closely related to feasibility is the notion of viability which implies feasibility. 
A specialized definition to account for the variable final time, terminal condition and the previously defined control parameterization is given as follows: % TODO: assumption hinzufügen? Oder viability ganz rausstreichen?
The tuple ($\rxset, \xfset$) is called \textit{viable for grid size $N$} if for each $\xs \in \rxset$ there exists $\tf \geq 0$ such that $\uadmset{\xs}{\tf} \neq \emptyset$ holds.
% The following assumption extends the notion of a viable set $\rxset$ from~\cite{gruene2002_springer} to incorporate $\xfset$ and the free final time $\tf$:
%\begin{assum}[Viability] \label{def:viability}
%	The tuple ($\rxset, \xfset$) is called \textit{viable} if for each $\xs \in \rxset$ there exists $\tf \geq t_0$ and $\uuc{t} \in \ruset$ such that $\xsolfun\big(t-t_0, \xs, \uuc{t} \big) \in \rxset$ and
%	$\xsolfun\big(\tf-t_0, \xs, \uuc{t} \big) \in \xfset$ hold for all $t \in [t_0, \tf]$.
%\end{assum}
%Viability is often also called weak or controlled forward invariance and provides a means to define controllability in the presence of state and control constraints. 

\subsection{Closed-Loop Control}

In the following, the previously defined \OCP{}~\eqref{eq:ocp} is integrated with state feedback. % TODO: magni2004_tac ~\cite{magni2004_tac, gruene2017_book}.
Since \eqref{eq:ocp} can only be solved at discrete time instances, the sampled feedback control law is defined according to the grid
$\tmud{0} < \tmud{1} < \dotsc < \tmud{n} < \dotsc < \infty$ with \mbox{$n \in \naturalzero$} and $\tmud{n} \in \realposzero$.
Hereby, subscript $\mu$ indicates that the context belongs to the evolution of the closed-loop system.
To account for the time-variable grid, the interval length $[\tmud{n}, \tmud{n+1})$ at time instance $n$ is inherited from the first interval of the corresponding prediction~\eqref{eq:ocp}.
In particular, the implicit control law $\fundef{\mulaw_N}{\xset}{\uset}$ for $\tmu \in [\tmud{n}, \tmud{n}+\vardtmud{n})$ with $\vardtmud{n} \mdef \vardtstar\big(\xmuc{\tmud{n}}, N\big)$ is defined by:
\begin{equation}
\mulaw_N\big(\xmuc{\tmu}\big) \mdef \uucstarbig{0,\xmuc{\tmud{n}}, N}. %\quad \text{for} \quad \tmu \in [\tmud{n}, \tmud{n}+\vardtstar\big(\xmuc{\tmu}\big). 
\label{eq:mulaw}
\end{equation}
Hereby, $\fundef{\xmu}{\realposzero}{\xset}$ denotes the closed-loop state trajectory which is either directly measured or obtained by a state observer.

Considering the plant dynamics~\eqref{eq:system_dynamics_cont} and control law~\eqref{eq:mulaw}, the resulting closed-loop system with initial state $\xmus \in \xset$ at time $\tmus$ is defined by:
\begin{equation}
\xmucdot{\tmu} = \f\Big(\xmuc{\tmu},\mulaw_N\big(\xmuc{\tmu}\big)\Big) , \quad \xmuc{\tmus}=\xmus. % TODO g ist schon nebenbedingung beim NLP!
\label{eq:closed_loop_system}
\end{equation}
According to~\eqref{eq:ivp:x}, the corresponding closed-loop state evolution is obtained by:
\begin{equation}
\xmuc{\tmu} \mdef \xmusolfun\big(\tmu,\tmus, \xmus \big) \mdef \xsolfun\Big(\tmu-\tmus, \xmus, \mulaw_N\big(\xmuc{\tmu}\big) \Big).
\label{eq:ivp:xmu}
\end{equation}

\section{Stability Analysis and Controller Design}
\label{sec:stability_analysis}

This section analyzes the stability properties of closed-loop system~\eqref{eq:closed_loop_system} and proposes an additional grid adaptation scheme to improve the closed-loop performance.  % TODO: so lassen?
Whereas standard \MPC{} formulations with terminal equality condition $\xfset\mdef \{ \xf \}$ usually enforce asymptotic stability~\cite{gruene2017_book},  % TODO andere quellen als nur grüne
this observation does not apply to time-variable grids and \OCP{}~\eqref{eq:ocp} in particular.
The following stability results explicitly account for these types of grids and are based on the principle of optimality~\cite{bertsekas1995_book}. %~\cite{bellmann1957}.
%framework of comparison functions and Lyapunov theory similar to standard \MPC{}. %~\cite{gruene2017_book}.
%Consequently, we first recapitulate some established definitions, c.f.~\cite{gruene2017_book}.
We first define the a controllability region specialized for this setup:

\begin{defn}[Controllability Region] \label{def:uniform:controllability_region}
	The set which contains all states $\xtilde \in \rxset$ from that the terminal set~$\xfset$ is reachable within $N$ steps and at least a transition time $t_c \in \realposzero$ is defined by:
	\begin{equation}
	\Pfull{N}{t_c} \mdef \big\{ \xtilde \in \rxset \mid \uadmset{\xtilde}{t} \neq \emptyset, 0 \leq t \leq t_c \big\}.  % TODO: Evtl. müssen wir das auf t = t_c fixieren?????
	\end{equation}
\end{defn}
Note, this set relates to viability up to time $t_c$. %  on $\big(\Pfull{N}{t_c}, \xfset\big)$
It is further equivalent to the reachable set from $\xfset$ in time $t_c$ and the backward respectively reverse-time system $\xcdot{t} = -\f\big(\xc{t},\uc{t}\big)$ \cite{blanchini2015_book}.
Determining $\Pfull{N}{t_c}$ analytically is usually difficult and common numerical methods to obtain reachable or controllable sets can be applied~\cite{blanchini2015_book}.
For instance simulations can be performed for low-dimensional systems, or an auxiliary optimal control problem can be solved which maximizes the target set (reachable set) w.r.t. the backward dynamics~\cite{horiuchi2015}. % TODO: andere literatur?
% TODO: dieser letzte teil evtl auch in den "Numerical Realization teil"

\subsection{Stability and Recursive Feasibility Issues} %

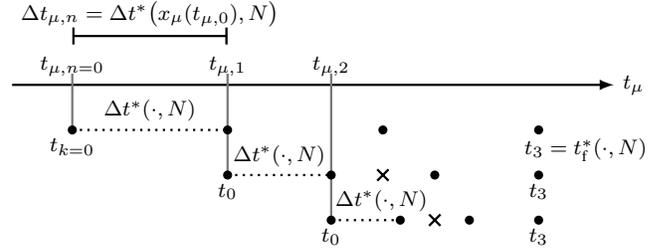
\begin{figure}[tb]
	\centering
	
	\tikzstyle{style1}=[black,solid,line width=1.25pt]%,mark=*,mark options={solid,fill=black,draw=black,scale=0.6}]
	\tikzstyle{style2}=[darkgray,solid,line width=1.25pt] %,mark=*,mark options={solid,fill=black,draw=black,scale=0.6}]
	\tikzstyle{style3}=[black,dashed,line width=1.25pt] %,mark=*,mark options={solid,fill=darkgray,draw=darkgray,scale=0.6}]
	\tikzstyle{style4}=[lightgray,solid,line width=1.25pt] %,mark=*,mark options={solid,fill=lightgray,draw=lightgray,scale=0.6}]
	\tikzstyle{style5}=[darkgray,dashed,line width=1.25pt] %,mark=*,mark options={solid,fill=lightgray,draw=lightgray,scale=0.6}]
	\tikzstyle{style6}=[lightgray,dashed,line width=1.25pt] %,mark=*,mark options={solid,fill=lightgray,draw=lightgray,scale=0.6}]
	\tikzstyle{style7}=[darkgray, dash pattern={on 7pt off 2pt on 1pt off 3pt},line width=1.25pt]%,mark=*,mark options={solid,fill=black,draw=black,scale=0.6}]
	
	\tikzstyle{timestep}=[draw, circle, fill, minimum size=0.09cm, inner sep=0pt]
	\tikzset{cross/.style={cross out, draw=black, minimum size=2*(#1-\pgflinewidth), inner sep=0pt, outer sep=0pt},
		%default radius will be 1pt. 
		cross/.default={1pt}}
	
	%	\tikzsetnextfilename{example_cl_async}%
	\begin{tikzpicture}[font=\footnotesize]
	\coordinate(origin) at (0.0,0);
	\draw[-latex, black] (origin) -- node[pos=1, right] {$\tmu$} ++(8.0,0);
	%\draw[-latex, darkgray] (origin) -- node[pos=1, left] {$\tmu$} ++(0,-1);
	
	\coordinate (t00) at (0.8, -0.6);
	\coordinate (t0f) at (7, -0.6);
	
	\coordinate (t01) at ($(t00)!0.333!(t0f)$) {};
	\coordinate (t02) at ($(t00)!0.666!(t0f)$) {};
	
	\coordinate (t10aux) at (1, -1.2);
	\coordinate (t10) at (t10aux -| t01);
	\coordinate (t1f) at (7, -1.2);
	
	\coordinate (t11) at ($(t10)!0.333!(t1f)$) {};
	\coordinate (t12) at ($(t10)!0.666!(t1f)$) {};
	
	\coordinate (t20aux) at (1, -1.8);
	\coordinate (t20) at (t20aux -| t11);
	\coordinate (t2f) at (7, -1.8);
	
	\coordinate (t21) at ($(t20)!0.333!(t2f)$) {};
	\coordinate (t22) at ($(t20)!0.666!(t2f)$) {};
	
	\draw[darkgray, shorten >=-0.15cm] (t00) -- node[pos=1, above] {\color{black}$\tmud{n=0}$} (origin-|t00);
	\draw[darkgray, shorten >=-0.15cm] (t10) -- node[pos=1, above] {\color{black}$\tmud{1}$} (origin-|t10);
	\draw[darkgray, shorten >=-0.15cm] (t20) -- node[pos=1, above] {\color{black}$\tmud{2}$} (origin-|t20);
	
	\node[timestep] at (t00) {};
	\node[timestep] at (t01) {};
	\node[timestep] at (t02) {};
	\node[timestep] at (t0f) {};
	
	\node[timestep] at (t10) {};
	\node[timestep] at (t11) {};
	\node[timestep] at (t12) {};
	\node[timestep] at (t1f) {};
	
	\node[timestep] at (t20) {};
	\node[timestep] at (t21) {};
	\node[timestep] at (t22) {};
	\node[timestep] at (t2f) {};
	
	\draw[dotted] (t00) -- node [pos=0.5, above] {$\vardtstar(\cdot,N)$} (t01);
	\draw[dotted] (t10) -- node [pos=0.5, above] {$\vardtstar(\cdot,N)$} (t11);
	\draw[dotted] (t20) -- node [pos=0.5, above, xshift=0.21cm] {$\vardtstar(\cdot,N)$} (t21);
	
	\node[below] at (t00) {$t_{k=0}$};
	\node[below] at (t10) {$t_{0}$};
	\node[below] at (t20) {$t_{0}$};
	\node[below right=0cm and -0.3cm] at (t0f) {$t_3=\tfstar(\cdot,N)$};
	\node[below] at (t1f) {$t_3$};
	\node[below] at (t2f) {$t_3$};
	
	\draw[|-|] ([yshift=1.25cm]t00) -- node[pos=0.5, above] {$\vardtmud{n}=\vardtstar\big(\xmuc{\tmud{0}}, N\big)$} ([yshift=1.25cm]t01);
	
	\draw (t11-|t02) node[cross=3pt] {};
	\draw (t21-|t12) node[cross=3pt] {};
	
	\end{tikzpicture}
	
	\caption{Illustration of the closed-loop grid with constant size~$N$. The dots correspond to the discretization grid of the prediction at closed-loop sampling times $\tmud{n}$.
		    Crosses indicate a potential lack of recursive feasibility.}
	\label{fig:control_samples_async}
\end{figure}

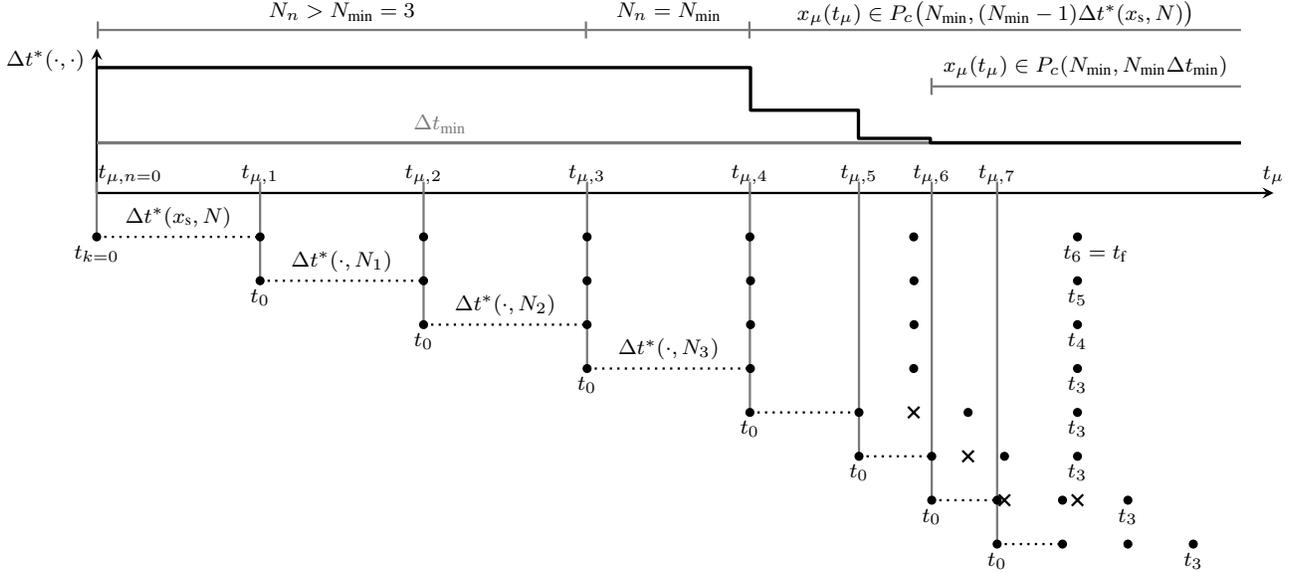
\begin{figure*}[!htbp]
	\centering
	%\tikzsetnextfilename{grid_resolution_feas_loss}%
	\input{practical_stab_illustration.tex}
	\caption{Illustration of the different stages during closed-loop control. The points correspond to the discretization grid of the prediction (horizon) at closed-loop sampling times $\tmud{n}$.
		Crosses indicate a potential lack of recursive feasibility for states within the region $\Pfullbig{\Nmin}{(\Nmin-1)\vardtstar(\xs,N)}$. As soon as $\vardtstar = \vardtmin$ is reached, that is for $\xmu(\tmu)\in \Pfull{\Nmin}{\Nmin\vardtmin}$, the cost no longer decreases and thus the temporal grid/horizon length is fixed and becomes receding. However, time-optimality and proper stabilization at $\xf$ is no longer guaranteed.}
	\label{fig:practical_stab_illustration}
\end{figure*}

% TODO: vielleicht wo anders
%The nonlinear programs in \refchap{chap:uniform_grid} contain the final state as an equality condition similar to the optimal control problems with time transformation for which stability and recursive feasibility are often \textit{assumed} to be valid.
Forward invariance is often ensured by maintaining recursive feasibility during closed-loop transition.
As a requirement, the first \OCP{} must be feasible which is, e.g., addressed by the viability assumption.
However, the control parameterization in \OCP{}~\eqref{eq:ocp} invalidates recursive feasibility guarantees for closed-loop system~\eqref{eq:closed_loop_system}. 
% these time-optimal control formulations do not necessarily realize a proper stabilization at the steady state.
%In detail, the reasons are as follows:
Since the grid is uniform with size $N$ and the final state is subject to terminal conditions, the time interval decreases as the closed-loop system evolves,  % TODO: hier könnte man kürzen, da Formel
i.e. $\vardtstar\big(\xmuc{\tmud{n}},N\big) \to \vardtmin$ for $n\to\infty$ (see \reffig{fig:control_samples_async}). % TODO: brauchen wir dafür ein PROOF?
Correspondingly, grid points $t_k$ of the very first \OCP{} do not coincide with the closed-loop sampling instances $\tmud{n}$.
Hence, Bellman's principle of optimality does not hold anymore. % TODO: Cite
The prediction at $\tmud{1}$ cannot realize a switch in control at $t_{2}$ of the previous solution (marked by a cross symbol).

In addition, $\vardtstar\big(\xmuc{\tmud{n}},N\big) \to 0$ for $n\to \infty$ and $\vardtmin=0$ implies the following drawbacks:
First, the terminal set/state~$\xfset$ cannot be reached in finite time.
Secondly and more technically, small time intervals result in ill-conditioned optimization problems and thus affect convergence.
% TODO erwähnen?
% By choosing a synchronous sampling time of the closed-loop system instead of inhering time intervals from the \OCP{}s would resolve these issues but recursive feasibiliy still cannot be guaranteed.

Consider the case $\xmuc{\tmu} \in \xfset$.
For $\vardtmin=0$, the optimal time interval is $\vardtstar(\xmuc{\tmu},N)= 0$ % TODO PROOF?
and hence evaluating control law~\eqref{eq:mulaw} reveals the following difficulties: %cannot be realized properly in practice.
Neither an infinite sampling rate can be realized in practice, nor does the imminent control action $\uucstar{0,\xmuc{\tmu},N}$ ensures keeping the system in $\xfset$.
As $\tf=0$ holds, any admissible control satisfies $\xsolfun\big(0,\xf,u(t)\big)$ in~\eqref{eq:ivp:x} and~\eqref{eq:ocp} and thus forward invariance of $\xfset$ cannot be guaranteed in general. % TODO: besser erklären?

%
%This in turn allows the optimizer to choose any $u(t) \in \ruset$ which is verified by $\xsolfun\big(0,\xf,u(t)\big)$ in~\eqref{eq:ivp:x} (underconstrained problem).
%Alternatively, the special case $\vardtstar=0$ must be handled separately when evaluating the sampled control law.
%%Consequently, the controls defining $u(t)$ are underconstrained and can be chosen arbitrarily leading to a displacement from the equilibrium.
%For $\vardtmin > 0$, $u(t)$ can also be arbitrary as long as the boundary conditions $\xmuc{\tmu} = \xf$ are met.
%Note, intermediate states are not attracted by $\xf$ as in quadratic form \MPC{} % TODO: verweis auf das lemma mit den optimalitätsbedingungen (oder der proof dazu)
%which often results in oscillations around the steady state and chattering in the control law while loosing feasibility and asymptotic stability guarantees. % TODO: bild?

% TODO: 
% Note that the previous list of issues also applies to approaches based on time transformation. % TODO: so nennen?
% TODO: hybrid cost time transformation?

\subsection{Grid Adaption and Closed-Loop Convergence}

This section proposes a modification that accounts for the previously discussed issues. % and analyzes
The lack of recursive feasibility and finite-time convergence are addressed by reducing the grid size $N$ while the closed-loop system evolves.
Let $N_n \in \naturalpos$ denote the grid size at time instance $\tmud{n}$ with $n \in \naturalzero$.  
The initial grid size $N_0 \mdef N$ is set to a user-defined $N$.
Subsequent grid sizes are then defined by $N_{n+1} \mdef \max(N_{n} - 1,\Nmin)$ for $n>0$.
%The purpose of $\Nmin\in\naturalpos$ is discussed later, first consider the case $\Nmin=1$.
Note that a minimum grid size $\Nmin \in \naturalpos$ is crucial to maintain viability.
In theory, if the first solution is feasible, all subsequent solutions are feasible for $N_n\to1$.
However, any small disturbance results in a potential loss of viability for small grid sizes in practice.
It is well known that systems often require multiple switches in control to reach $\xf$, even for initial states close to $\xf$.
For unconstrained linear systems, \cite{kailath1980_book, vandenbroeck2011_ifac} suggest to choose $\Nmin \geq p / q$ with state dimension $p$ and control dimension~$q$ respectively. 
A proper value for $\Nmin$ depends on the system and constraint sets, however, choosing some $\Nmin > p$ is a good starting point for simulations and experiments. 
% TODO: remarks?? assumptions?

Taking the grid adaption into account, control law~\eqref{eq:mulaw} for $\tmu \in [\tmud{n}, \tmud{n}+\vardtmud{n})$  with $\vardtmud{n} \mdef \vardtstar\big(\xmuc{\tmud{n}}, N_n\big)$ results in:
% TODO: evtl können wir nicht \mulaw_{N_n} verwenden, weil das nicht zur closed-loop system beschriebung passt. Subscript N bzw N_n sollte vielleicht weg!
\begin{equation}
\mulaw\big(\xmuc{\tmu}\big) \mdef \uucstarbig{0,\xmuc{\tmud{n}}, {N_n}}. %\quad \text{for} \quad \tmu \in [\tmud{n}, \tmud{n}+\vardtstar\big(\xmuc{\tmu}\big). 
\label{eq:mulaw_adapt}
\end{equation}
% TODO: verwenden?
% This in turn ensures recursive feasibility at least until $\Nmin=1$ is reached (see \reffig{fig:cl:adapt}). 
%
In addition, choosing $\vardtmin > 0$ appropriately circumvents the numerical ill-conditioning and zero interval lengths $\vardtmud{n}=0$ for states inside $\xfset$.
However, as soon as $\vardtstar(\xmu,N)=\vardtmin$ for some $\xmu$ and $N$ is reached, i.e. for $\xmu \in \Pfull{N}{N\vardtmin}$, the cost in \eqref{eq:ocp} remains constant which in turn affects closed-loop convergence.

% The terminal equality constraint requires the system to perfectly reach $\xf$ with a single control command ($\Nmin=1$).

Figure~\ref{fig:practical_stab_illustration} shows the different stages during closed-loop control depending on $\Nmin$ and $\vardtmin$.
As long as $N_n>\Nmin$ lasts, the closed-loop evolution coincides with the initially predicted trajectories.
Afterward, as soon as $N_n=\Nmin$ is reached, the controller takes one more step with $\vardtmud{n}=\vardtstar(\xs,N)$ before the temporal resolution increases and hence loosing recursive feasibility guarantees.
For the active lower bound $\vardtstar(\cdot,N_n)=\vardtmin$ (third stage), the cost function remains constant and hence asymptotic stability can no longer be maintained. 
%In particular, the principle of optimality does not hold anymore as the cost function no longer constitutes a suitable Lyapunov function, and further convergence to $\xf$ cannot be ensured. % TODO: weitere Ausführungen???
%
These observations are captured by the following results:
%The following lemmas provide intermediate results which are later required to verify Lyapunov function candidates.

\begin{lem}
	Let $\tfstar(\xs,N) \in \realpos$ denote the optimal transition time obtained from~\eqref{eq:ocp} with $\xs \in \rxset$,  $\xfset = \{ \xf \}$ and grid size $N\geq 1$.
	Further assume that the solution is feasible. % TODO: schöner formulieren?!
	Then, relation
	\begin{equation}
	\tfstar(\xs,N) > N\vardtmin \iff \xs \notin \Pfull{N}{N\vardtmin}
	\label{app:eq:vardt_not_in_p}
	\end{equation}
	holds for $\vardtmin \geq 0$.
	%The corresponding state trajectory $\xuc{t}$  obtained from either~\eqref{eq:uniform:fd:ocp}, \eqref{eq:uniform:ms:ocp} or~\eqref{eq:uniform:hs:ocp} with  $\xfset = \{ \xf \}$.
	%Further assume that \refassump{assum:uniform:optimal_solution} and \refassump{assump:ode_conditions} hold.
	\label{lemma:const_cost_dtstar_dtmin}
\end{lem}
\begin{IEEEproof}
	First, we abbreviate $\tfstar\mdef\tfstar(\xs,N)$, define $\tfmin\mdef N\vardtmin$ and consider the case $\xs \notin \Pfull{N}{\tfmin} \implies \tfstar > \tfmin$.
	The implication follows immediately from the definition of $\Pfull{N}{\tfmin}$ even for non-optimal $\tf$.
	By contraposition, the implication is equivalent to $\tfstar \leq \tfmin \implies \xs \in \Pfull{N}{\tfmin}$.
	The optimal solution is feasible by assumption and hence $\uadmset{\xs}{\tfstar} \neq \emptyset$ and $\tfstar \geq \tfmin$ are ensured such that condition $\tfstar \leq \tfmin$ is replaced by $\tfstar = \tfmin$.
	Consequently, all requirements for $\xs \in \Pfull{N}{\tfmin}$ are met. % TODO: ist das hier zu viel blabla?
	
	The second case $\tfstar > \tfmin \implies \xs \notin \Pfull{N}{\tfmin}$ does not hold for arbitrary (non-optimal) $\tf$ since control trajectories $u \in \uadmset{\xs}{\tf}$ could exists which start and end in $\Pfull{N}{\tfmin}$ but fulfill $\tfstar > \tfmin$ (for example keeping the system at the steady state).
	However, to show that the implication holds for $\tfstar$ subject to~\eqref{eq:ocp},
	consider the contraposition $\xs \in \Pfull{N}{\tfmin} \implies \tfstar \leq \tfmin$. 
	If $\xs \in \Pfull{N}{\tfmin}$ holds, then $\uadmset{\xs}{t}\neq \emptyset$ for $0 \leq t \leq \tfmin$ by \refdef{def:uniform:controllability_region}.
	Solving~\eqref{eq:ocp} results in minimum-time solutions adhering to constraint \mbox{$\tfmin \leq \tfstar$}
	and hence the only feasible transition time for $\xs \in \Pfull{N}{\tfmin}$ is $\tfstar=\tfmin$.
	The existence of this particular $u \in \mathscr{U}^N$ is confirmed by assumption (feasibility). % even though it does not need to be unique.
	Consequently, $\tfstar=\tfmin$ proves the original implication $\xs \in \Pfull{N}{\tfmin} \implies$ \mbox{$\tfstar \leq \tfmin$}. 
	
	Finally, equivalence~\eqref{app:eq:vardt_not_in_p} follows immediately since both implications are true.
	%\qed	
	%since $\Pxf{N}$  all 
\end{IEEEproof}

\begin{prop} \label{shrinking_horizon_mpc:p_stability}
	Consider system~\eqref{eq:system_dynamics_cont} with initial state \mbox{$\xs\in\rxset$}, final state $\xfset = \{ \xf \}$, optimal control problem~\eqref{eq:ocp} and control law~\eqref{eq:mulaw_adapt}. % with piecewise constant controls
	%Let $\xfset = \{ \xf \}$ represent a steady state such that there exists $u \in \ruset$ with $\f \big(\xf,u\big) = 0$.
	Choose $\Nmin\geq 1$ and define $\vardtmin\geq 0$ and $N\geq \Nmin$ ensuring $\vardtmin < \vardtstar(\xs,N)$. % TODO: hier schon feasibility assum?
	Furthermore, assume that the initial solution to~\eqref{eq:ocp} is feasible. % TODO: extra assumption?
	Then, the closed-loop system reaches the region $P=\Pxfshrinking$.
\end{prop}
% TODO: the set P is defined due to potential loss of recursive feasibility. Otherwise, the lyapunov function still decreases towards Pxf{Nmin}$
%Refer to \refapp{app:extensions:uniform:stability} for the proof.
\begin{IEEEproof}
	The proof relies on the dynamic programming principle~\cite{bertsekas1995_book} and hence its mathematical exposition is kept brief.
	Let $P$ abbreviate $P\mdef \Pxfshrinking$, $\xmu$ the current state $\xmu \mdef \xmuc{\tmud{n}}$ and $\xmu^+$ the successor state $\xmu^+ \mdef  \xmusolfun\big(\tmud{n+1},\tmud{n}, \xmu(\tmud{n}) \big)$. % \xmu(\tmud{n+1}) \mdef
	%
	%As for \MPC{} in general, $V$ is chosen as the optimal cost function value.	
	% TODO: Since \refprop{shrinking_horizon_mpc:p_stability} demands to choose $0 \leq \vardtmin < \vardtstar(\xs,N)$ for the first \OCP{}, $\tfstar(\xs,N) > N\vardtmin$ implies $\xmu \notin \Pfull{N}{N\vardtmin}$ according to \reflem{lemma:const_cost_dtstar_dtmin}.
	%	
	The solution to the first \OCP{} at time $\tmud{0}$ is feasible by assumption and hence an admissible control trajectory $u \in \uadmset{\xs}{\tfstar(\xs,N)}$ exists. % The optimal time interval is $\vardtstar$.
	%	To emphasize its direct relation to the optimal control problem at time instance $\tmud{n}$, the optimal time interval is now denoted by $\vardtstar.
	Note that the grid is adapted with $N_0\mdef N$ and $N_{n+1} \mdef \max(N_{n} - 1,\Nmin)$.	
	First, consider the case $N_n > \Nmin$.
	The optimal cost function value is $V(\xmu) = \tfstar(\xmu, N_n)$.
	Applying the principle of optimality results in
	\begin{align}
		\tfstar(\xmu,N_n) &= \vardtstar(\xmu,N_n) + (N_n-1) \vardtstar(\xmu,N_n) \nonumber \\
		&= \vardtstar(\xmu,N_n) + (N_n-1) \vardtstar(\xmu^+,N_n-1) \nonumber \\
		\Leftrightarrow  V(\xmu) &= \vardtstar(\xmu,N_n) + V(\xmu^+).
		\label{eq:dyn_prog_time_optimal}
	\end{align}
	Consequently, the solution at $\xmu^+$ with grid size $N_n-1$ coincides with the previous solution, and hence the nominal closed-loop and open-loop evolution are identical.
	This includes recursive feasibility which implies forward invariance as well as the control to $P$.
	It can be easily verified that \eqref{eq:dyn_prog_time_optimal} only holds as long as $N_n$ can be reduced by one in each step. % which holds for all states $\xmu \notin P$.  
	$N_n\geq\Nmin>0$ is ensured by definition.   
	As soon as $N_n=\Nmin$ is reached, control law~\eqref{eq:mulaw_adapt} performs one more step with $\vardtstar(\xs,N_n)$.
	Afterward, $\vardtstar(\xmu,\Nmin)$ decreases in each step (see \reffig{fig:practical_stab_illustration}) so it does not match the optimal cost value in the previous step, invalidating~\eqref{eq:dyn_prog_time_optimal}.
	%
	%As stated previously, $\xmu \notin \Pfull{N_n}{N\vardtmin}$ follows from \reflem{lemma:const_cost_dtstar_dtmin} and $\vardtstar(\xs,N_n) > \vardtmin$ as $\vardtstar(\xmu,N_n)$ remains constant for all $N_n > \Nmin$ (see~\eqref{eq:dyn_prog_time_optimal}). % TODO: letzteres vielleicht noch ausführlicher einbauen, wir droppen das hier einfach
	%
	%Note, the principle of optimality~\eqref{eq:dyn_prog_time_optimal} implies recursive feasibility and hence forward invariance on~$\rxset$ for $N_n > \Nmin$.
	%As soon as $N=\Nmin$ is reached, control law~\eqref{eq:mulaw_adapt} performs one more step with $\vardtstar(\xs,N)$.
	%Afterwards, $\vardtstar(\xmu,\Nmin)$ decreases in each step (see \reffig{fig:practical_stab_illustration}) until the lower bound $\vardtstar(\xmu,\Nmin)\geq\vardtmin$ becomes active.
	%This in turn leads to a potential lack of recursive feasibility and hence forward invariance cannot be ensured in $P$ % TODO: extra lemma für forward invariance, und das P die richtige beschreibung ist?
	%which limits the stability results to $\rxset \setminus P$.	
	%\qed
\end{IEEEproof}

%The closed-loop system looses recursive feasibility guarantees in $P$ and then asymptotic stability in $\Pfull{\Nmin}{\Nmin\vardtmin}$.
Condition $\vardtmin < \vardtstar(\xs,N)$ ensures that the closed-loop state is not inside $\Pfull{\Nmin}{\Nmin\vardtmin}$ before the systems enters $P$ and applies for proper choices of $\vardtmin$ and $N$ (\reflem{lemma:const_cost_dtstar_dtmin}). % TODO: referenz auf lemma im Anhang! TODO: dies ist eigentlich optional
Note, \refprop{shrinking_horizon_mpc:p_stability} even holds for $\vardtmin \leq \vardtstar(\xs,N)$, however, we exclude this case to avoid ambiguous non-time optimal solutions as described before.
%Note that the result allows for selecting $\vardtmin=0$ even though not realizable in praxis. % TODO; dieser satz weg?
%
For arbitrary systems, \refprop{shrinking_horizon_mpc:p_stability} does not guarantee true practical stability, i.e. that the state is ultimately bounded to $P$ after arrival. 

Practical implications of these results are that by increasing the initial $N$, which in turn reduces $\vardtstar(\cdot,\cdot)$, or by decreasing $\Nmin$, the size of region~$P$ is reduced.
In contrast, certain choices of $N$ and $\Nmin$ are further implicitly bounded by $\vardtmin$, viability and the computational resources, thus limiting the endless decrease of~$P$.
However, the specific control applications then decide if this particular region is small enough to, for example, realize a proper point-to-point motion.
Alternatively, it facilitates the systematic design of a dual-mode controller as described in \refsec{sec:dual-mode}.

%It is particularly important to still provide stability and feasibility guarantees even for $\Nmin>1$.
%In the following, the stability results are relaxed to ensure convergence at least up to a certain region containing $\xf$.

% TODO: folgende section oder Remark???!
% \section{Extended Grid Adaptation}
\begin{rem} \label{rem:grid_adaptation}
	Recursive feasibility does not generally hold for adaptation schemes that adjust the grid resolution w.r.t. a desired sample time $\vardtref \in \realpos$ as in \cite{roesmann2015_ecc}.
	However, these schemes are particularly interesting in applications for which viability is assumed for the whole state space and for changes in $\xf$. 
	An extended adaptation strategy with hysteresis \mbox{$\vardthyst\in \realposzero$} is given as follows:
	\begin{align*}
	\vardtstar_n &\mdef \vardtstar\big(\xmu(\tmud{n}),N_n\big), \\
	N_{n+1} &=
	\begin{dcases}
	\max \big( \Phi(\vardtstar_n,N_n), \Nmin \big) & |\vardtstar_n - \vardtref| > \vardthyst \\
	\max (N_{n}-1,\Nmin)       & \mtext{otherwise}
	\end{dcases}.
	%\label{eq:uniform:adapt}
	\end{align*}
	 Temporal adaption in $\Phi(\vardtstar_n,N_n)$ could be achieved by estimation, i.e. $\Phi(\vardtstar_n,N_n) = \min (  N_n \vardtstar_n / \vardtref , \Nmax ) $ and $\Nmax \in \naturalpos$ as safeguard,
	 or by linear search $\Phi(\vardtstar_n,N_n) =  N_n+1$ for $\vardtstar_n > \vardtref + \vardthyst$ resp. $\Phi(\vardtstar_n,N_n) =  N_n-1$ for $\vardtstar_n < \vardtref - \vardthyst$.
	 By setting $\vardtref = \vardtstar(\xs,N)$, adaptation is inactive and hence convergence holds according to~\refprop{shrinking_horizon_mpc:p_stability} and since $|\vardtstar_n - \vardtref| \leq \vardthyst$ holds for all $n$ outside~$P$.
\end{rem}

\section{Direct Transcription}
\label{sec:numerical_realization}

\subsection{Formulation}

This section addresses the realization of \OCP{}~\eqref{eq:ocp} in terms of two different nonlinear program formulations
that retain the inherent sparse structure of standard \MPC{}.
%
% Afterward, a dedicated nominal stability result 
The first formulation is referred to as global uniform grid approach.
Let $\vardt$, $\u_k$ with $k=0,1,\dotsc,N-1$ and $\x_k$ with $k=0,1,\dotsc,N$ denote the parameters subject to optimization.
%Note that $\vardt$ replaces $\tf/N$ for better readability.
Accordingly, the nonlinear program  is defined as follows:
\begin{gather}
	%\tag{CTOCP}
	\underset{ \substack{ \uud{0}, \uud{1}, \dotsc, \uud{N-1}, \\ \xud{0}, \xud{1}, \dotsc, \xud{N}, \\  \vardt} }{\min}\ N\vardt
	\label{eq:global_uniform_nlp} \\
	\hspace{-6.8cm}\text{subject to} \nonumber \\ 
	\begin{align*}
		&\xud{0} = \xs, \quad \xud{N} \in \xfset,\quad \xud{k} \in \rxset, \quad \uud{k} \in \ruset,\\ %,\quad  t_n=T\\
		&\vardtmin \leq \vardt \leq \vardtmax,\quad \xud{k+1}  =  \xsolfun( \vardt, \xud{k}, \uud{k} ), \\
		&k=0,1,\dotsc,N-1.
	\end{align*}
\end{gather}
Local optimization solvers often assume that $\rxset, \ruset$ and $\xfset$ are compact and convex.
In every practical implementation these sets are replaced by algebraic equality and inequality constraint functions which is usually straightforward and not described here in detail.
Note that the sparsity pattern of the Hessian of Lagrangian, e.g., contains a single dense row and column for the parameter~$\vardt$.
Another nonlinear program that is larger, though sparser, results from the definition of individual $\vardt_k$ and a uniformity condition $\vardt_k = \vardt_{k+1}$:
\begin{gather}
	%\tag{CTOCP}
	\underset{ \substack{ \uud{0}, \uud{1}, \dotsc, \uud{N-1}, \\ \xud{0}, \xud{1}, \dotsc, \xud{N}, \\  \vardt_0, \vardt_1, \dotsc, \vardt_{N-1}} }{\min}\ \sum_{k=0}^{N-1}\vardt_k
	\label{eq:local_uniform_nlp} \\
	\hspace{-6.8cm}\text{subject to} \nonumber \\ 
	\begin{align*}
		&\xud{0} = \xs, \quad \xud{N} \in \xfset,\quad \xud{k} \in \rxset, \quad \uud{k} \in \ruset,\\ %,\quad  t_n=T\\
		&\vardtmin \leq \vardt_0 \leq \vardtmax,\quad \vardt_k = \vardt_{k+1},\\ 
		& \xud{k+1}  =  \xsolfun( \vardt_k, \xud{k}, \uud{k} ),\quad k=0,1,\dotsc,N-1.
	\end{align*}
\end{gather}
This formulation is referred to as local uniform grid approach.
% TODO: oder vllt doch?:
% A performance comparison between~\eqref{eq:global_uniform_nlp} and~\eqref{eq:local_uniform_nlp} is not in the scope of this paper. 
\begin{prop} \label{prop:numerical_ocps}
	The solutions to \OCP{}s~\eqref{eq:ocp}, \eqref{eq:global_uniform_nlp} and~\eqref{eq:local_uniform_nlp} are identical.
\end{prop}
\begin{IEEEproof} % TODO: vielleicht ist das hier auch einfach too much und kann in einem satz gesagt werden! -> Trivial!!
	Even if the proof is mostly trivial, we include it for the sake of completeness. % TODO lassen?
	First, we show that \OCP{}s~\eqref{eq:ocp} and~\eqref{eq:global_uniform_nlp} coincide and so does their solution.
	The control function space $\uadmset{\xs}{\tf}\subseteq\upcset{\tf}$ is defined according to a uniform grid with size $N$ and partition length $\vardt$.
	Accordingly, $\tf = N\vardt$ defines the cost function in~\eqref{eq:global_uniform_nlp}.
	As the control trajectory is piecewise constant w.r.t. the grid, i.e. $\u(t) \mdef \u_k \mtext{ for } t \in [t_k, t_k+\vardt)$, it is completely described by parameters $\u_k$ with $k=0,1,\dotsc,N-1$ and $\vardt$.
	To account for the admissibility conditions in $\uadmset{\xs}{\tf}$, let $\x_{k} \mdef \xsolfun\big(k\tf /N,\x_s,\u(t)\big)$ denote the states at grid points $t_k$.
	According to~\eqref{eq:ivp:x}, it is $\x_0 = \xs$ and by time-invariance of~\eqref{eq:system_dynamics_cont} $\x_{k+1} = \xsolfun\big(\vardt,\x_k,\u_k\big) \mtext{ for } k=0,1,\dotsc,N-1$.
	Control constraints in $\uadmset{\xs}{\tf}$ are imposed by enforcing $u_k \in \ruset$ and state constraints at grid points by $\x_k \in \rxset$ for $k=0,1,\dotsc,N-1$.
	The last state must adhere to $\x_N = \xsolfun\big(\tf,\x_s,\u(t)\big) \in \xfset$. By uniformity, $N\vardtmin \leq \tf \leq N\vardtmax$ is substituted by $\vardtmin \leq \vardt \leq \vardtmax$.
	Consequently, minimizing~\eqref{eq:global_uniform_nlp} w.r.t. all $u_k$, all $x_k$ and $\vardt$ leads to the same solution as~\eqref{eq:ocp}. % TODO: kann man das wirklich so lassen?
	
	Showing that the solutions to \eqref{eq:global_uniform_nlp} and~\eqref{eq:local_uniform_nlp} coincide is straightforward. Equality constraint $\vardt_k = \vardt_{k+1}$ in~\eqref{eq:local_uniform_nlp} ensures uniformity for the optimal solution
	such that $\vardtstar_k(\xs,N) = \vardtstar(\xs,N)$ holds for all $k=0,1,\dotsc,N-1$ and hence the minimum cost is $\sum_{k=0}^{N-1} \vardtstar_k(\xs,N) = N \vardtstar(\xs,N)$ which coincides with~\eqref{eq:global_uniform_nlp}.
	The same applies to constraints. % TODO ausführlich genug?
%	\qed
\end{IEEEproof}
Necessary and sufficient optimality conditions for general nonlinear programs apply~\cite{nocedal2006_book}.
Note that the deflection constraint $\xud{k+1}  =  \xsolfun( \vardt_k, \xud{k}, \uud{k} )$ in~\eqref{eq:global_uniform_nlp} and~\eqref{eq:local_uniform_nlp} is continuously differentiable w.r.t. $\vardt_k, \xud{k}, \uud{k}$ and $\xud{k+1}$ even though 
the grid is temporally variable. % TODO: PROOF?
Any practical realization solves the initial value problem~\eqref{eq:ivp:x} numerically, e.g. by one-step methods (Euler, Runge-Kutta) that maintain continuous differentiability.
Hereby, the theory of sampled-data systems applies which usually requires fast sampling~\cite{nesic2004_tac}. 
By choosing the Euler family or the implicit trapezoidal rule, i.e. $\xsolfun(\vardt_k,\xud{k}, \uud{k}) \approx \xud{k} + \vardt_k \f(\xud{k}, \uud{k})$ for forward Euler, an interesting nominal asymptotic stability result follows under certain conditions:
%Nevertheless, nominal asymptotic stability for $\xfset\mdef\{\xf\}$ holds under certain conditions:
\begin{prop} \label{shrinking_horizon_mpc:asymp_stability}
	Consider system~\eqref{eq:system_dynamics_cont} with initial state $\xs\in\rxset$, \OCP{}~\eqref{eq:global_uniform_nlp} or~\eqref{eq:local_uniform_nlp}, control law~\eqref{eq:mulaw_adapt} and Euler resp. trapezoidal integration. % with piecewise constant controls
	%Assume that $N$ is chosen sufficiently large such that the dynamics error is negligible (fast sampling~\cite{nesic2004_tac}).
	Let $\xfset = \{ \xf \}$ represent a steady state such that there exists $u \in \ruset$ with $\f \big(\xf,u\big) = 0$.
	Choose $\Nmin=1$ and define $\vardtmin>0$ and $N \geq 1$ ensuring $\vardtmin < \vardtstar(\xs,N)$. % TODO: hier schon feasibility assum?
	Furthermore, assume that the initial solution is feasible and that constraint qualifications as well as second-order necessary conditions hold. % TODO: extra assumption?
	Then, the closed-loop system is asymptotically stable on~$\rxset$.
\end{prop}
\begin{IEEEproof}
	The proof is based on \refprop{shrinking_horizon_mpc:p_stability}.
	Since the minimum grid size is set to $\Nmin=1$, the principle of optimality ensures that the system actually reaches $P = \Pxfshrinking = \xf$. 
	%	$\xmu(\tmud{n}) \in \rxset\setminus P$ with $P\mdef \Pxfshrinking$. 
	%Note, in this case it is $P=\{\xf\}$.
	%	Furthermore, the dynamic programming principle ensures that the system actually reaches $\xf$:
	As soon as the grid reduces to $N=\Nmin$ for some $\xmu(\tmud{n})$ the successor state is $\xmu(\tmud{n+1})=\xf$.
	%The Lyapunov conditions are ensured up to only $\xmu(\tmud{n}) \in \Pfull{\Nmin}{\tfmin}$ (\reflem{lemma:cost_lyapunov_bounds}).
	Due to $\Nmin=1$ and condition $\vardtmin < \vardtstar(\xs,N)$, the system reaches $\xf$ before the optimal time interval reduces to $\vardtmin$.
	Since $\vardtmin>0$,  the Karush–Kuhn–Tucker conditions ensure recursive feasibility and nominal asymptotic stability beyond $\xf$:
	The set $\ruset$ is replaced by an algebraic description in any practical realization (see \refsec{sec:numerical_realization}).
	Let this set defined by $g(\uud{0}) \leq 0$ with $\fundef{g}{\uset}{\mathbb{R}^S}$. 
	State restrictions $\rxset$ are fulfilled implicitly since $N=1$ and initial respectively final states are fixed to $\xf$.
	
	We show now that first-order optimality conditions (see~\cite{nocedal2006_book}) ensure to find $\uudstar{0}\in \ruset$ such that $\f \big(\xf,\uudstar{0}\big) = 0$ holds.
	Since $N=1$, states are directly substituted and the remaining optimization parameters are $\uud{0}$ and $\vardt_k = \vardt = \tf$.
	%Accordingly, further state constraints are omitted.
	Applying the forward Euler method to~\eqref{eq:ivp:x} results in the equality constraint \mbox{$h(\uud{0},\vardt) \mdef \xf- \xf + \vardt \f(\xf,\uud{0}) = 0$}.
	
	The Lagrangian with multipliers $\lambda_0 \in \mathbb{R}^p$, $\mu_0 \in \mathbb{R}^S$ and $\mu_1 \in \mathbb{R}$ is given by
	%\begin{equation}
	%\begin{split}
	$	\mathcal{L}(\uud{0}, \vardt, \lambda_0, \mu_0, \mu_1) =
	\vardt + \lambda_0^\transpose h(\uud{0},\vardt) + \mu_0^\transpose g(\uud{0}) + \mu_1(\vardtmin-\vardt)$.
	%\end{split}
	%	\end{equation}
	The first-order optimality conditions for the optimal parameters (indicated by a star) are:
	\begin{subequations}
		\label{kkt:fd}
		\begin{align}
			\gradwrt{\mathcal{L}}{\uud{0}}(\cdot) &= \gradwrt{\big( \lambda_0^{*\transpose}  h(\uudstar{0},\tf)\big)}{\uud{0}}\\
			&+ \gradwrt{\big( \mu_0^{*\transpose} g(\uudstar{0}) \big)}{\uud{0}}  = \nullvec, \label{kkt:fd:1} \\ 
			\gradwrt{\mathcal{L}}{\vardt}(\cdot)  &= 1 + \lambda_0^\transpose \f(\xf,\uudstar{0}) - \mu^*_1 = \nullvec, \label{kkt:fd:2} \\
			h(\uudstar{0},\vardtstar) &= \nullvec, \label{kkt:fd:3} \\
			\mu_0^{*\transpose} g(\uudstar{0}) &= \nullvec, \label{kkt:fd:4} \\
			\mu^*_1(\vardtmin-\vardtstar) &= \nullvec, \label{kkt:fd:5}\\
			g(\uudstar{0}) \leq \nullvec, &\ \vardtstar \geq \vardtmin, \ \mu^*_0 \geq \nullvec, \ \mu^*_1 \geq 0. % TODO: \vardtmax!
			%		\gradwrt{\mathcal{L}}{z}(z^*, \lambda^*, \mu^*) = \nullvec,  &  \\
			%		h_i(z^*) = \nullvec,  &\quad \textup{for all } i \in \mathcal{E}, \\
			%		g_i(z^*) \leq \nullvec,  &\quad \textup{for all } i \in \mathcal{I}, \\
			%		\mu^*_i \geq \nullvec,  &\quad \textup{for all } i \in \mathcal{I}, \\
			%		\mu^*_i g_i(z^*) = \nullvec,  &\quad \textup{for all } i \in \mathcal{I}.
		\end{align} 
	\end{subequations}
	Parameter $\vardtstar \geq \vardtmin > 0$ is strictly positive by definition.
	Consequently, \eqref{kkt:fd:3} ensures that $\uudstar{0}$ satisfies $f(\xf,\uudstar{0})=0$ and that particular $\uudstar{0}$ exists by definition and $g(\uudstar{0}) \leq \nullvec$ holds.
	%Combining \eqref{kkt:fd:2}, \eqref{kkt:fd:5} and $f(\xf,\uudstar{0})=0$ confirms that the optimal time interval $\vardtstar$ satisfies $\vardtstar=\vardtmin$.
	Combining \eqref{kkt:fd:2} and $f(\xf,\uudstar{0})=0$ implies $\mu^*_1=1$ which in turn confirms with \eqref{kkt:fd:5} that the optimal time interval $\vardtstar$ satisfies $\vardtstar=\vardtmin$.
	Equations~\eqref{kkt:fd:1} and~\eqref{kkt:fd:4} are fulfilled by proper choices of $\lambda^*_0$ and $\mu^*_0$.
	For example, if control constraints $g(\cdot)$ are inactive, $\mu^*_0=\lambda^*_0=\nullvec$ are possible solutions.
	For the other one-step methods, i.e. backward Euler and trapezoidal rule, $h(\uudstar{0},\vardtstar) = \nullvec$ also immediately implies $f(\xf,\uudstar{0})=0$ which is skipped for brevity.
	%Hereby, the deflection constraint in~\eqref{eq:global_uniform_nlp} or~\eqref{eq:local_uniform_nlp} is substituted by $\xud{k+1}  =  \xsolfun( \vardt_k, \xud{k}, \uud{k} )$
	
	%Note, most nonlinear program solvers determine solutions according to the first-order optimality conditions.
	Constraint qualification and second-order sufficient conditions for a true feasible (local) minimizer are ensured by assumption. % (see~\cite{nocedal2006_book}).
	%These are assumed to hold for any particular system dynamics equation.
	%\qed
\end{IEEEproof}
% TODO: initial solution is feasible follows for example from viability
% TODO: hier auf die necessary conditions usw eingehen?
The result inherently addresses recursive feasibility and hence ensures forward invariance on $\rxset$. % according to \refdef{def:asymptotic_stability}.
%\refappc{app:extensions:uniform:stability} provides the proof and related lemmas. % lemmata?
Satisfying condition $\vardtmin < \vardtstar$ is straightforward by a proper choice of~$N$. % TODO proofs vielleicht im ANhang noch in mehrere sections splitten
However, the result is rather theoretical as $\Nmin=1$ is often required for robustness in practice (refer to the discussion in~\refsec{sec:stability_analysis}).
%It is important to notice, that the proof requires piecewise constant controls. 
%Note, this result already takes~$\vardtmin$ into account according to \refrem{remark:dt_min} and condition $\vardtmin < \vardtstar$ is easily satisfied by a proper choice of~$N$.

% TODO: diss wortlaut
%\refpropc{shrinking_horizon_mpc:asymp_stability} is a rather theoretical result.

\subsection{Hypergraph Representation}

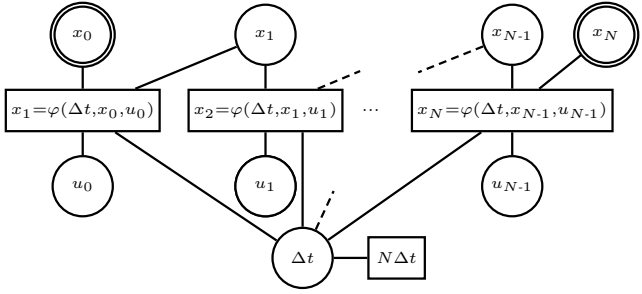
\begin{figure}[tb]
	\centering
	%\tikzsetnextfilename{hypergraph_to_full_discretization_global}%
	\input{hypergraph_full_discretization_global.tex}
	\caption{Hypergraph of the global uniform grid}
	% TODO: bild schön machen (verbindungen stimmen noch nicht alle in der mitte)
	\label{fig:hypergraph:to:fd:global}
\end{figure}
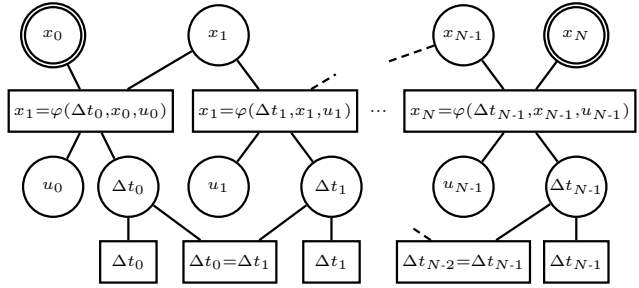
\begin{figure}[tb]
	\centering
%	\tikzsetnextfilename{hypergraph_to_full_discretization_local}%
	\input{hypergraph_full_discretization_local.tex}
	\caption{Hypergraph of the local uniform grid}
	\label{fig:hypergraph:to:fd:local}
\end{figure}

\begin{figure}[tb]
	\centering
	\begin{subfigure}{.4\columnwidth}
		\centering
		%\tikzsetnextfilename{hypergraph_sparsity_vdp_fd_jac}%
		\input{vdp_fd_hlag.tex}
		\caption{Global uniform grid} 
		\label{fig:hypergraph:sparsity:fd_hlag}
	\end{subfigure}\quad \quad
	\begin{subfigure}{.4\columnwidth}
		\centering
		%\tikzsetnextfilename{hypergraph_sparsity_vdp_fe_jac}%
		\input{vdp_nu_fd_hlag.tex}
		\caption{Local uniform grid} 
		\label{fig:hypergraph:sparsity:nu_fd_hlag}
	\end{subfigure}
	\caption{Structure of the Hessian of the Lagrangian for a second order system with $N=10,p=2$ and $q=1$.}
	\label{fig:hypergraph:sparsity_jac_fdms}
\end{figure}
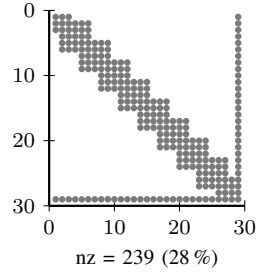
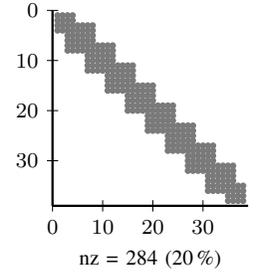

The local and the global uniform grid reveal different sparsity patterns due to their optimization structure.
This is also immediately visible in its hypergraph representation as introduced for MPC in~\cite{roesmann2018_aim}.
A hypergraph is a graph composed of a set of vertices and a set of hyperedges.
Hyperedges connect an arbitrary number of vertices rather than only pairs of vertices compared to regular graphs.
For nonlinear programs~\eqref{eq:global_uniform_nlp} and~\eqref{eq:local_uniform_nlp}, each vertex refers to an optimization parameter, i.e. $\xud{k}$, $\uud{k}$ or $\vardt_k$.
A hyperedge refers to cost or constraint terms and is only connected to the nodes on which they directly depend.
\reffigc{fig:hypergraph:to:fd:global} shows the hypergraph for the global uniform grid \eqref{eq:global_uniform_nlp}.
Trivial substitutions for optimization parameters like $\xud{0}=\xs$ are not represented by a dedicated edge as these parameters are not subject to optimization. 
These vertices are called fixed which is indicated by a double circle. % in \reffig{fig:hypergraph:to:fd:global}.
Also lower and upper bounds for optimization parameters are directly cached in their vertices.
Note that parameter $\vardt$ in \reffig{fig:hypergraph:to:fd:global} is connected with all edges for adhering to the system dynamics.
This indicates a dense column in the constraint Jacobian respectively a dense row and column in the Hessian of the Lagrangian (see \reffig{fig:hypergraph:sparsity:fd_hlag}).
In contrast, the hypergraph for the local uniform grid \eqref{eq:local_uniform_nlp} contains more vertices,
but the maximum number of connected vertices for each edge is limited and independent of $N$. 
\reffigc{fig:hypergraph:sparsity:nu_fd_hlag} shows an example for the corresponding structure of the Hessian of the Lagrangian in which the percentage of non-zeros (nz) is smaller.

The hypergraph representation is well suited for the practical implementation. 
It allows efficient computations of derivatives based on sparse finite differences. 
The graph eliminates the need for an extra graph coloring algorithm to find the sparsity patterns.
Block Jacobian and Hessian matrices are straightforward to calculate by iterating edges or vertices in the graph.
%The computational burden for the solution of the nonlinear programs increases almost linearly in relation to the grid size. 
% TODO -> auswertung
Furthermore, the grid size adaptation as described in \refsec{sec:stability_analysis} leads to the ongoing change of problem dimensions.
Reconfiguring the hypergraph online while maintaining the inherent sparse structure requires almost negligible overhead which is crucial for real-time control.
Refer to~\cite{roesmann2018_aim} for a detailed description and general performance results.
%A performance comparison between the particular global and local uniform grids is not in the scope of this paper. 
%However, for small to mid size problems the computation times are comparable to each other and in both cases increase almost linearly in relation to the grid size. % TODO: WIRKLICH?

\section{Quasi-Time-Optimal Dual-Mode Control} % TODO: wo soll die bessere grid adaption hin?
\label{sec:dual-mode}

%The previous examples show that time-optimal control leads to chattering while stabilizing the system at the desired steady state.
%In practice, often chattering drastically limits the lifetime of mechanical components and hence a smooth stabilizing control is generally preferred.
%Two approaches are presented below but the exposition is kept brief since established state feedback control and conventional \MPC{} concepts apply.

As mentioned before, the results summarized in \refprop{shrinking_horizon_mpc:p_stability} facilitate the dual-mode controller design.
Dual-mode \MPC{} is the predecessor to quasi-infinite horizon \MPC{} concerning stability enforcement~\cite{mayne2000_automatica}. 
The key idea is to control the system to a terminal region $\rxsetlin$ and then switch to an external local stabilizing controller~\cite{chisci1996_ejoc,michalska1993_tac}.
In addition, stabilization with the dual-mode realization does not suffer from extensive chattering like actual time-optimal controllers do. % Formulierung

% TODO: thesis wortlaut
Let $\xf$ denote the steady state at which the system should be stabilized and $\uref$ the corresponding control such that $\f \big(\xf, \uref\big) = 0$ holds.
Assume that the linearized system $\xlindot(t) = A_c \xlin(t) + B_c \ulin(t)$ with $A_c \mdef \jacobwrt{\f}{\x}(\xeq, \uref)$ and $B_c \mdef \jacobwrt{\f}{\u}(\xeq, \uref)$ is stabilizable. % TODO: notation, entweder nach vorne, oder hier erlären (Jacobian)
Then, a linear state feedback $K \in \mathbb{R}^{q\times p}$ can be determined such that $A_c-B_c K$ is asymptotically stable.
However, since the original plant is nonlinear and constraints are present, the region of operation is limited to some set $\rxsetlin \subseteq \rxset$.
The set $\rxsetlin$ is determined such that the local feedback law $\mulawlin\big(\xmuc{\tmu}\big) = K (\xf - \xmuc{\tmu}) + \uref$
is admissible with $\mulawlin\big(\xmu\big) \in \ruset$ for all $\xmu \in \rxsetlin$.
Furthermore, the closed-loop system must be rendered forward invariant, in particular
%$\varp\big(\xmu,\mulawlin(\xmu)\big) \in \rxsetlin$ 
$\xsolfun\big(\tmu, \xmus, \mulaw\big(\xmuc{\tmu}\big) \big) \in \rxsetlin$
for all  $\xmus \in \rxsetlin$ and $\tmu\geq0$.  % TODO: diskret, continuierlich?!
These requirements on $\rxsetlin$ ensure feasibility.
However, to also ensure asymptotic stability, $\rxsetlin$ is further limited to provide a sufficient decrease of a suited Lyapunov function for the nonlinear closed-loop system. 
For a profound description refer to \cite{chen1982_automatica,michalska1993_tac}.
% TODO: in den example teil:
%\begin{rm}
%	The linear feedback controller is usually obtained by the well known linear quadratic regulator (LQR) design. % TODO: wir die abkürzung später verwendet?
%	In practice, it is often sufficient to determine $\rxsetlin$ by discretizing the state space in the vicinity of $\xf$ and by performing closed-loop simulations with the linear controller and the nonlinear system.
%\end{rm} % TODO: für den LQR ist die lyapunov function klar

Combining the control law from the linear controller design above with~\eqref{eq:mulaw_adapt} results in the following quasi-time-optimal dual-mode control law:
\begin{equation}
\label{eq:dual_mode_law}
\mulawdual\big(\xmuc{\tmu}\big) =
\begin{cases}
\mulaw\big(\xmuc{\tmu}\big) & \mtext{for } \xmuc{\tmu} \notin $\rxsetlin$ \\
\mulawlin\big(\xmuc{\tmu}\big) & \mtext{for } \xmuc{\tmu} \in $\rxsetlin$
\end{cases}.
\end{equation}
\begin{rem}
	The sampling rate of $\mulawlin$ is subject to the local controller design. A practical realization might also consider a discrete-time LQR (see \refsec{sec:example}).
\end{rem}
In order to ensure feasibility of the combined control law, $\xfset \subseteq \rxsetlin$ must hold, since otherwise the time-optimal controller would not 
reach $\rxsetlin$. 
%In fact, the theoretical investigations in \refsec{sec:uniform:shrinking_horizon_cl} considered $\xfset = \{ \xf \}$ as stated in \refthm{shrinking_horizon_mpc:p_stability}.
Accordingly, even $\Pxfshrinking \subseteq \rxsetlin$ must hold to ensure a proper convergence to~$\rxsetlin$:
\begin{cor} \label{dual_mode_stability_result}
	Consider the closed-loop control system and the assumptions according to \refprop{shrinking_horizon_mpc:p_stability} with design parameters $N, \Nmin$ and $\vardtmin$.
	The resulting control law is denoted by $\mulaw(\xmu)$. % TODO: vllt. \funde{}{}{} verwenden...
	Furthermore, consider that a control law $\mulawlin(\xmu)$ exists which ensures asymptotic stability and forward invariance for the nonlinear system for all states $\xmu \in \rxsetlin$.
	Then, the closed-loop control system with composite control law~\eqref{eq:dual_mode_law} and steady state~$\xf$ is asymptotically stable on~$\rxset$ if $\Pxfshrinking \subseteq \rxsetlin$ holds. % TODO:ist das system oder die ruhelage asymptotisch stabil?
\end{cor}
% TODO: Hinweis auf proof im ANhang. sollte trivial sein
%The proof is straightforward and follows immediately from the previous results (refer to \refapp{app:extensions:uniform:stability}).
\begin{IEEEproof}
	The closed-loop system according to \refprop{shrinking_horizon_mpc:p_stability} converges to $\Pxfshrinking$ by ensuring recursive feasibility. 
	By assumption, the local linear controller ensures forward invariance and asymptotic stability on $\rxsetlin$.
	Since control law~\eqref{eq:dual_mode_law} switches to the local controller as soon as~$\rxsetlin$ is reached,
	condition $\Pxfshrinking \subseteq \rxsetlin$ immediately ensures that $\xf$ is asymptotically stable for the whole dual-mode control system.
	%\qed
\end{IEEEproof}

% TODO: discussion: die results brauchen leider das ergebnis der ersten ocp lösung -> vardtstar!

%% file: practical_stab_illustration.tex
\tikzstyle{style1}=[black,solid,line width=1.25pt]%,mark=*,mark options={solid,fill=black,draw=black,scale=0.6}]
\tikzstyle{style2}=[darkgray,solid,line width=1.25pt] %,mark=*,mark options={solid,fill=black,draw=black,scale=0.6}]
\tikzstyle{style3}=[black,dashed,line width=1.25pt] %,mark=*,mark options={solid,fill=darkgray,draw=darkgray,scale=0.6}]
\tikzstyle{style4}=[lightgray,solid,line width=1.25pt] %,mark=*,mark options={solid,fill=lightgray,draw=lightgray,scale=0.6}]
\tikzstyle{style5}=[darkgray,dashed,line width=1.25pt] %,mark=*,mark options={solid,fill=lightgray,draw=lightgray,scale=0.6}]
\tikzstyle{style6}=[lightgray,dashed,line width=1.25pt] %,mark=*,mark options={solid,fill=lightgray,draw=lightgray,scale=0.6}]
\tikzstyle{style7}=[darkgray, dash pattern={on 7pt off 2pt on 1pt off 3pt},line width=1.25pt]%,mark=*,mark options={solid,fill=black,draw=black,scale=0.6}]

\tikzstyle{timestep}=[draw, circle, fill, minimum size=0.09cm, inner sep=0pt]
\tikzset{cross/.style={cross out, draw=black, minimum size=2*(#1-\pgflinewidth), inner sep=0pt, outer sep=0pt},
	%default radius will be 1pt. 
	cross/.default={1pt}}

\begin{tikzpicture}[font=\footnotesize]

\setlength{\figurewidth}{0.95\linewidth}
\setlength{\figureheight}{3.5cm}

\begin{axis}[
width=\figurewidth,
height=\figureheight,
xmin=0,
xmax=7.2,
xlabel={$\tmu$},
xtick=\empty,
xticklabels={,,}, % hide x tick labels
x label style={at={(axis description cs:1.0,0)},anchor=south},
ymin=0,
ymax=1.15,
ytick=\empty,
ylabel={$\vardtstar(\cdot,\cdot)$},
yticklabels={,,}, % hide x tick labels
y label style={at={(axis description cs:-0.003,0.92)},anchor=east,rotate=-90},
every tick/.style={darkgray},
%every axis label/.append style ={darkgray},
%every tick label/.append style={darkgray},  
clip=false, 
]
\coordinate(origin) at (0.0,0);
%\draw[-latex, darkgray] (origin) -- node[pos=1, right] {$\tmu$} ++(4.5,0);
%\draw[-latex, darkgray] (origin) -- node[pos=1, left] {$\tmu$} ++(0,-1);

\coordinate (t00) at (0, -0.35);
\coordinate (t0f) at (6, -0.35);

\coordinate (t01) at ($(t00)!0.1666!(t0f)$) {};
\coordinate (t02) at ($(t00)!0.3333!(t0f)$) {};
\coordinate (t03) at ($(t00)!0.5!(t0f)$) {};
\coordinate (t04) at ($(t00)!0.666!(t0f)$) {};
\coordinate (t05) at ($(t00)!0.833!(t0f)$) {};

\coordinate (t10aux) at (1, -0.7);
\coordinate (t10) at (t10aux -| t01);
\coordinate (t1f) at (6, -0.7);

\coordinate (t11) at ($(t10)!0.2!(t1f)$) {};
\coordinate (t12) at ($(t10)!0.4!(t1f)$) {};
\coordinate (t13) at ($(t10)!0.6!(t1f)$) {};
\coordinate (t14) at ($(t10)!0.8!(t1f)$) {};

\coordinate (t20aux) at (1, -1.05);
\coordinate (t20) at (t20aux -| t02);
\coordinate (t2f) at (6, -1.05);

\coordinate (t21) at ($(t20)!0.25!(t2f)$) {};
\coordinate (t22) at ($(t20)!0.5!(t2f)$) {};
\coordinate (t23) at ($(t20)!0.75!(t2f)$) {};

\coordinate (t30aux) at (1, -1.4);
\coordinate (t30) at (t30aux -| t03);
\coordinate (t3f) at (6, -1.4);

\coordinate (t31) at ($(t30)!0.333!(t3f)$) {};
\coordinate (t32) at ($(t30)!0.666!(t3f)$) {};
%\coordinate (t33) at ($(t30)!0.75!(t3f)$) {};

\coordinate (t40aux) at (1, -1.75);
\coordinate (t40) at (t40aux -| t04);
\coordinate (t4f) at (6, -1.75);

\coordinate (t41) at ($(t40)!0.333!(t4f)$) {};
\coordinate (t42) at ($(t40)!0.666!(t4f)$) {};

\coordinate (t50aux) at (1, -2.1);
\coordinate (t50) at (t50aux -| t41);
\coordinate (t5f) at (6, -2.1);

\coordinate (t51) at ($(t50)!0.333!(t5f)$) {};
\coordinate (t52) at ($(t50)!0.666!(t5f)$) {};

\coordinate (t60aux) at (1, -2.45);
\coordinate (t60) at (t60aux -| t51);

\coordinate (t61) at ($(t60)+(0.4,0)$) {};
\coordinate (t62) at ($(t60)+(0.8,0)$) {};
\coordinate (t6f) at ($(t60)+(1.2,0)$) {};

\coordinate (t70aux) at (1, -2.8);
\coordinate (t70) at (t70aux -| t61);

\coordinate (t71) at ($(t70)+(0.4,0)$) {};
\coordinate (t72) at ($(t70)+(0.8,0)$) {};
\coordinate (t7f) at ($(t70)+(1.2,0)$) {};

\draw[darkgray, shorten >=-0.15cm] (t00) -- node[pos=1, above right=0cm and -0.1cm] {\color{black}$\tmud{n=0}$} (origin-|t00);
\draw[darkgray, shorten >=-0.15cm] (t10) -- node[pos=1, above] {\color{black}$\tmud{1}$} (origin-|t10);
\draw[darkgray, shorten >=-0.15cm] (t20) -- node[pos=1, above] {\color{black}$\tmud{2}$} (origin-|t20);
\draw[darkgray, shorten >=-0.15cm] (t30) -- node[pos=1, above] {\color{black}$\tmud{3}$} (origin-|t30);
\draw[darkgray, shorten >=-0.15cm] (t40) -- node[pos=1, above] {\color{black}$\tmud{4}$} (origin-|t40);
\draw[darkgray, shorten >=-0.15cm] (t50) -- node[pos=1, above] {\color{black}$\tmud{5}$} (origin-|t50);
\draw[darkgray, shorten >=-0.15cm] (t60) -- node[pos=1, above] {\color{black}$\tmud{6}$} (origin-|t60);
\draw[darkgray, shorten >=-0.15cm] (t70) -- node[pos=1, above] {\color{black}$\tmud{7}$} (origin-|t70);

\node[timestep] at (t00) {};
\node[timestep] at (t01) {};
\node[timestep] at (t02) {};
\node[timestep] at (t03) {};
\node[timestep] at (t04) {};
\node[timestep] at (t05) {};
\node[timestep] at (t0f) {};

\node[timestep] at (t10) {};
\node[timestep] at (t11) {};
\node[timestep] at (t12) {};
\node[timestep] at (t13) {};
\node[timestep] at (t14) {};
\node[timestep] at (t1f) {};

\node[timestep] at (t20) {};
\node[timestep] at (t21) {};
\node[timestep] at (t22) {};
\node[timestep] at (t23) {};
\node[timestep] at (t2f) {};

\node[timestep] at (t30) {};
\node[timestep] at (t31) {};
\node[timestep] at (t32) {};
\node[timestep] at (t3f) {};

\node[timestep] at (t40) {};
\node[timestep] at (t41) {};
\node[timestep] at (t42) {};
\node[timestep] at (t4f) {};

\node[timestep] at (t50) {};
\node[timestep] at (t51) {};
\node[timestep] at (t52) {};
\node[timestep] at (t5f) {};

\node[timestep] at (t60) {};
\node[timestep] at (t61) {};
\node[timestep] at (t62) {};
\node[timestep] at (t6f) {};

\node[timestep] at (t70) {};
\node[timestep] at (t71) {};
\node[timestep] at (t72) {};
\node[timestep] at (t7f) {};

\draw[dotted] (t00) -- node [pos=0.5, above] {$\vardtstar(\xs,N)$} (t01); % pos=0.5, above=-0.1cm
\draw[dotted] (t10) -- node [pos=0.5, above] {$\vardtstar(\cdot,N_1)$} (t11);
\draw[dotted] (t20) -- node [pos=0.5, above] {$\vardtstar(\cdot,N_2)$} (t21);
\draw[dotted] (t30) -- node [pos=0.5, above] {$\vardtstar(\cdot,N_3)$} (t31);
\draw[dotted] (t40) -- (t41); % node [pos=0.5, above] {$\vardtstar(\cdot,N_4)$} 
\draw[dotted] (t50) -- (t51); % node [pos=0.5, above] {$\vardtstar$}
\draw[dotted] (t60) -- (t61); % node [pos=0.5, above] {$\vardtstar$}
\draw[dotted] (t70) -- (t71); % node [pos=0.5, above] {$\vardtstar$}

\node[below] at (t00) {$t_{k=0}$};
\node[below] at (t10) {$t_{0}$};
\node[below] at (t20) {$t_{0}$};
\node[below] at (t30) {$t_{0}$};
\node[below] at (t40) {$t_{0}$};
\node[below] at (t50) {$t_{0}$};
\node[below] at (t60) {$t_{0}$};
\node[below] at (t70) {$t_{0}$};
\node[below right=0cm and -0.3cm] at (t0f) {$t_6=\tf$};
\node[below] at (t1f) {$t_5$};
\node[below] at (t2f) {$t_4$};
\node[below] at (t3f) {$t_3$};
\node[below] at (t4f) {$t_3$};
\node[below] at (t5f) {$t_3$};
\node[below] at (t6f) {$t_3$};
\node[below] at (t7f) {$t_3$};

%	\draw[|-|] ([yshift=1.25cm]t00) -- node[pos=0.5, above, xshift=0.7cm] {$\vardtmud{n}=\vardtstar=\mtext{const.}$} ([yshift=1.25cm]t01);
%\draw[|-|] ([yshift=1.25cm]t00) -- node[pos=0.5, above] {$\vardtmud{n}=\vardtstar$} ([yshift=1.25cm]t01);

\draw[style2] (0, 0.4) -- node[pos=0.3, above] {$\vardtmin$} (7, 0.4);
\addplot [style1, const plot] coordinates {(0,1.0) (4,0.66) (4.66,0.4356) (5.1,0.4) (7,0.4)};

% draw lack of feasibility
\draw (t41-|t05) node[cross=3pt] {};
\draw (t51-|t42) node[cross=3pt] {};
\draw (t61-|t52) node[cross=3pt] {};
\draw (t61-|t5f) node[cross=3pt] {};

% draw sections

\draw[|-|, darkgray] (0,1.3) -- node [pos=0.5, above, black] {$N_n > \Nmin = 3$} (3,1.3);
\draw[-|, darkgray] (3,1.3) -- node [pos=0.5, above, black] {$N_n = \Nmin$} (4,1.3);
\draw[-, darkgray] (4,1.3) -- node [pos=0.5, above=-0.08cm, black] {$\xmu(\tmu) \in \Pfullbig{\Nmin}{(\Nmin-1)\vardtstar(\xs,N)}$} (7,1.3);

\draw[|-, darkgray] (5.1,0.85) -- node [pos=0.5, above, black] {$\xmu(\tmu) \in \Pfull{\Nmin}{\Nmin\vardtmin}$} (7,0.85);

\end{axis}
\end{tikzpicture}

%% file: hypergraph_full_discretization_global.tex
\begin{tikzpicture}[font=\notsotiny, node distance=1.0 cm]

\newcommand{\shortminus}{\text{-}}

% reduce horizontal spacing in math mode (https://tex.stackexchange.com/questions/41913/how-to-get-less-spacing-in-math-mode)
\setmuskip{\thinmuskip}{0mu plus 0mu}
\setmuskip{\medmuskip}{0mu plus 0mu} 
\setmuskip{\thickmuskip}{0mu plus 0mu}

\tikzstyle{vertex}=[black, draw, circle, inner sep = 0pt, minimum size = 0.8cm] % TODO unify all hypergraph figures
\tikzstyle{fixed_vertex}=[vertex, double]
\tikzstyle{edge}=[black, draw, rectangle, inner sep = 2pt, minimum width = 0.75cm, minimum height = 0.55cm]

% Vertices
\node [vertex] (u0) at (0,0) {$\uud{0}$};
\coordinate [vertex, right= 1.6cm of u0] (u1);
%\node[right=1.5cm of u1] (udots) {$\cdots$};
\node[right=0.875cm of u1] (udots) {};
\node [vertex, right= 1.35cm of udots] (uf) {$\uud{N\shortminus1}$};

\node [fixed_vertex, above= 1.6cm of u0.center] (x0) {$\xud{0}$};
\node [vertex] (x1) at (x0 -| u1) {$\xud{1}$};
\coordinate [right = 2.0cm of x1] (x2); % just for the dashed line
%\node[] at (x0 -| udots)  (xdots) {$\cdots$};
\node[] at (x0 -| udots)  (xdots) {};
\node [vertex, above= 1.6cm of uf.center] (xNm1) {$\xud{N\shortminus1}$};
\node [fixed_vertex, xshift=1.25cm] at (xNm1)  (xf)  {$\xud{N}$};

\coordinate (xaux) at ($(u0)!0.42!(xf)$);
\node [vertex, yshift=-1.8cm] at (xaux) (dt) {$\vardt$};

% Edges
\coordinate (yaux) at ($(u0)!0.5!(x0)$);
%\node[edge, anchor=center, align=center] at (yaux -| u0) (edge0) {$h_f\big(\xud{0}, \uud{0}, \vardt, \xud{1}\big)$};
\node[edge, anchor=center, align=center] at (yaux -| u0) (edge0) {$\xud{1} = \xsolfun( \vardt, \xud{0}, \uud{0} )$};

\coordinate (yaux) at ($(u1)!0.5!(x1)$);
%\node[edge, anchor=center, align=center] at (yaux -| u1) (edge1) {$h_f\big(\xud{1}, \uud{1}, \vardt, \xud{2}\big)$};
\node[edge, anchor=center, align=center] at (yaux -| u1) (edge1) {$\xud{2} = \xsolfun( \vardt, \xud{1}, \uud{1} )$};

\coordinate (yaux) at ($(uf)!0.5!(xNm1)$);
%\node[edge, anchor=center, align=center] at (yaux -| uf) (edge2) {$h_f\big(\xud{N-1}, \uud{N-1}, \vardt, \xud{N}\big)$};
\node[edge, anchor=center, align=center] at (yaux -| uf) (edge2) {$\xud{N} = \xsolfun( \vardt, \xud{N\shortminus1}, \uud{N\shortminus1} )$};

\node[edge, anchor=center, align=center, xshift=1.25cm] at (dt) (edgedt) {$N\vardt$};

%\node[edge, anchor=south, align=center] at ($(uf.north)!0.5!(xf.south)$) (edgeN) {$V_f(\mathbf{x}_N)$;\\$\mathbf{x}_N \in \mathbb{X}_f$};

\node[] at ($(udots)!0.5!(xdots)$) (edgedots) {$\cdots$};

\draw (u0) -- (edge0);
\draw (dt) -- (edge0);
\draw (x0) -- (edge0);
\draw (x1) -- (edge0);

\draw (u1) -- (edge1);
\coordinate (edge1_right) at ($(edge1.south)!0.6!(edge1.south east)$);
\draw (dt) -- (edge1_right -| dt);
\draw (x1) -- (edge1);
%\draw (x2) -- (edge1);
\draw [densely dashed, shorten >=1.25cm] (edge1) -- (x2);

\draw [densely dashed] (xNm1) -- ++(-1.25cm,-0.5cm);

\draw (uf) -- (edge2);
\draw (dt) -- (edge2);
\draw (xNm1) -- (edge2);
\draw (xf) -- (edge2);

\draw [densely dashed, shorten >=1.0cm] (dt) -- (edgedots);

\draw (dt) -- (edgedt);

% draw u1 in front of the lines
\node [vertex, fill=white] at (u1) {$\uud{1}$};

\end{tikzpicture}%

%% file: hypergraph_full_discretization_local.tex
\begin{tikzpicture}[font=\notsotiny, node distance=2.2 cm]

\newcommand{\shortminus}{\text{-}}

% reduce horizontal spacing in math mode (https://tex.stackexchange.com/questions/41913/how-to-get-less-spacing-in-math-mode)
\setmuskip{\thinmuskip}{0mu plus 0mu}
\setmuskip{\medmuskip}{0mu plus 0mu} 
\setmuskip{\thickmuskip}{0mu plus 0mu}

\tikzstyle{vertex}=[black, draw, circle, inner sep = 0pt, minimum size = 0.8cm] % TODO unify all hypergraph figures
\tikzstyle{fixed_vertex}=[vertex, double]
\tikzstyle{edge}=[black, draw, rectangle, inner sep = 2pt, minimum width = 0.75cm, minimum height = 0.55cm]

% Vertices
\node [vertex] (u0) at (0,0) {$\uud{0}$};
\node [vertex, xshift=1.0cm] at (u0) (dt0) {$\vardt_0$};
\node [vertex, right of= u0] (u1) {$\uud{1}$};
\node [vertex, xshift=1.5cm] at (u1) (dt1) {$\vardt_1$};
%\node[right=1.5cm of u1] (udots) {$\cdots$};
\node[right=1.6cm of u1] (udots) {};
\node [vertex, right= 0.6cm of udots] (uf) {$\uud{N\shortminus1}$};
\node [vertex, xshift=1.5cm] at (uf) (dtf) {$\vardt_{N\shortminus1}$};

\node [fixed_vertex, above= 1.6cm of u0.center] (x0) {$\xud{0}$};
\node [vertex] (x1) at (x0 -| u1) {$\xud{1}$};
\coordinate [right= 2.0cm of x1] (x2); % just for the dashed line
%\node[] at (x0 -| udots)  (xdots) {$\cdots$};
\node[] at (x0 -| udots)  (xdots) {};
\node [vertex, above= 1.6cm of uf.center] (xNm1) {$\xud{N\shortminus1}$};
\node [fixed_vertex, xshift=0.0cm] (xf) at (x0 -| dtf) {$\xud{N}$};

% Edges
\coordinate (xaux) at ($(u0)!0.5!(dt0)$);
\coordinate (yaux) at ($(u0)!0.5!(x0)$);
%\node[edge, anchor=center, align=center] at (yaux -| xaux) (edge0) {$h_f\big(\xud{0}, \uud{0}, \vardt_0, \xud{1}\big)$};
\node[edge, anchor=center, align=center] at (yaux -| xaux) (edge0) {$\xud{1} = \xsolfun( \vardt_0, \xud{0}, \uud{0} )$};

\coordinate (xaux) at ($(u1)!0.5!(dt1)$);
\coordinate (yaux) at ($(u1)!0.5!(x1)$);
%\node[edge, anchor=center, align=center] at (yaux -| xaux) (edge1) {$h_f\big(\xud{1}, \uud{1}, \vardt_1, \xud{2}\big)$};
\node[edge, anchor=center, align=center] at (yaux -| xaux) (edge1) {$\xud{1} = \xsolfun( \vardt_1, \xud{1}, \uud{1} )$};

\coordinate (xaux) at ($(uf)!0.5!(dtf)$);
\coordinate (yaux) at ($(uf)!0.5!(xNm1)$);
%\node[edge, anchor=center, align=center] at (yaux -| xaux) (edge2) {$h_f\big(\xud{N-1}, \uud{N-1}, \vardt_{N-1}, \xud{N}\big)$};
\node[edge, anchor=center, align=center] at (yaux -| xaux) (edge2) {$\xud{N} = \xsolfun( \vardt_{N\shortminus1}, \xud{N\shortminus1}, \uud{N\shortminus1} )$};

\node[edge, anchor=center, align=center, yshift=-1.0cm] at (dt0) (edgedt0) {$\vardt_0$};
\node[edge, anchor=center, align=center, yshift=-1.0cm] at (dt1) (edgedt1) {$\vardt_1$};
\node[edge, anchor=center, align=center, yshift=-1.0cm] at (dtf) (edgedt2) {$\vardt_ {N\shortminus1}$};

\node[edge, anchor=center, align=center, yshift=-1.0cm] at ($(dt0)!0.5!(dt1)$) (edgedt1dt2) {$\vardt_0 = \vardt_1$};
\coordinate [xshift=-1.5cm] (dtfn2) at (uf);
\node[edge, anchor=center, align=center, yshift=-1.0cm] at ($(dtfn2)!0.5!(dtf)$) (edgedtn2dtf) {$\vardt_{N\shortminus2}= \vardt_{N\shortminus1}$};

%\node[edge, anchor=south, align=center] at ($(uf.north)!0.5!(xf.south)$) (edgeN) {$V_f(\mathbf{x}_N)$;\\$\mathbf{x}_N \in \mathbb{X}_f$};

\node[] at ($(udots)!0.5!(xdots)$) (edgedots) {$\cdots$};

\draw (u0) -- (edge0);
\draw (dt0) -- (edge0);
\draw (x0) -- (edge0);
\draw (x1) -- (edge0);

\draw (u1) -- (edge1);
\draw (dt1) -- (edge1);
\draw (x1) -- (edge1);
%\draw (x2) -- (edge1);
\draw [densely dashed, shorten >=1.0cm] (edge1) -- (x2);

\draw [densely dashed] (xNm1) -- ++(-1.0cm,-0.35cm);

\draw (uf) -- (edge2);
\draw (dtf) -- (edge2);
\draw (xNm1) -- (edge2);
\draw (xf) -- (edge2);

\draw (dt0) -- (edgedt0);
\draw (dt1) -- (edgedt1);
\draw (dtf) -- (edgedt2);

\draw (dt0) -- (edgedt1dt2);
\draw (dt1) -- (edgedt1dt2);

\draw (dtf) -- (edgedtn2dtf);
\draw [densely dashed, shorten >=1.0cm]  (edgedtn2dtf) -- (dtfn2);

\end{tikzpicture}%

%% file: vdp_fd_hlag.tex
\definecolor{mycolor1}{rgb}{0.00000,0.44700,0.74100}%
\setlength{\figurewidth}{2.6cm}
\setlength{\figureheight}{2.6cm}

\begin{tikzpicture}[font=\footnotesize] % TODO font size?!

\begin{axis}[%
width=\figurewidth,
height=\figureheight,
at={(1.648in,0.642in)},
scale only axis,
xmin=0,
xmax=30,
x axis line style=-, % remove arrow tips
y axis line style=-, % remove arrow tips
xlabel={nz = 239 (\SI{28}{\%})},
y dir=reverse,
ymin=0,
ymax=30,
axis background/.style={fill=white}
]
\addplot [color=darkgray, draw=none, mark size=0.8pt, mark=*, mark options={solid, darkgray}, forget plot]
  table[]{
  1	1
  1	2
  1	3
  1	29
  2	1
  2	2
  2	3
  2	4
  2	5
  2	6
  2	29
  3	1
  3	2
  3	3
  3	4
  3	5
  3	6
  3	29
  4	2
  4	3
  4	4
  4	5
  4	6
  4	29
  5	2
  5	3
  5	4
  5	5
  5	6
  5	7
  5	8
  5	9
  5	29
  6	2
  6	3
  6	4
  6	5
  6	6
  6	7
  6	8
  6	9
  6	29
  7	5
  7	6
  7	7
  7	8
  7	9
  7	29
  8	5
  8	6
  8	7
  8	8
  8	9
  8	10
  8	11
  8	12
  8	29
  9	5
  9	6
  9	7
  9	8
  9	9
  9	10
  9	11
  9	12
  9	29
  10	8
  10	9
  10	10
  10	11
  10	12
  10	29
  11	8
  11	9
  11	10
  11	11
  11	12
  11	13
  11	14
  11	15
  11	29
  12	8
  12	9
  12	10
  12	11
  12	12
  12	13
  12	14
  12	15
  12	29
  13	11
  13	12
  13	13
  13	14
  13	15
  13	29
  14	11
  14	12
  14	13
  14	14
  14	15
  14	16
  14	17
  14	18
  14	29
  15	11
  15	12
  15	13
  15	14
  15	15
  15	16
  15	17
  15	18
  15	29
  16	14
  16	15
  16	16
  16	17
  16	18
  16	29
  17	14
  17	15
  17	16
  17	17
  17	18
  17	19
  17	20
  17	21
  17	29
  18	14
  18	15
  18	16
  18	17
  18	18
  18	19
  18	20
  18	21
  18	29
  19	17
  19	18
  19	19
  19	20
  19	21
  19	29
  20	17
  20	18
  20	19
  20	20
  20	21
  20	22
  20	23
  20	24
  20	29
  21	17
  21	18
  21	19
  21	20
  21	21
  21	22
  21	23
  21	24
  21	29
  22	20
  22	21
  22	22
  22	23
  22	24
  22	29
  23	20
  23	21
  23	22
  23	23
  23	24
  23	25
  23	26
  23	27
  23	29
  24	20
  24	21
  24	22
  24	23
  24	24
  24	25
  24	26
  24	27
  24	29
  25	23
  25	24
  25	25
  25	26
  25	27
  25	29
  26	23
  26	24
  26	25
  26	26
  26	27
  26	28
  26	29
  27	23
  27	24
  27	25
  27	26
  27	27
  27	28
  27	29
  28	26
  28	27
  28	28
  28	29
  29	1
  29	2
  29	3
  29	4
  29	5
  29	6
  29	7
  29	8
  29	9
  29	10
  29	11
  29	12
  29	13
  29	14
  29	15
  29	16
  29	17
  29	18
  29	19
  29	20
  29	21
  29	22
  29	23
  29	24
  29	25
  29	26
  29	27
  29	28
  29	29
};
\end{axis}
\end{tikzpicture}%

%% file: vdp_nu_fd_hlag.tex
\definecolor{mycolor1}{rgb}{0.00000,0.44700,0.74100}%
\setlength{\figurewidth}{2.6cm}
\setlength{\figureheight}{2.6cm}

\begin{tikzpicture}[font=\footnotesize] % TODO font size?!

\begin{axis}[%
width=\figurewidth,
height=\figureheight,
at={(1.648in,0.642in)},
scale only axis,
xmin=0,
xmax=39,
x axis line style=-, % remove arrow tips
y axis line style=-, % remove arrow tips
xlabel={nz = 284 (\SI{20}{\%})},
y dir=reverse,
ymin=0,
ymax=39,
axis background/.style={fill=white}
]
\addplot [color=darkgray, draw=none, mark size=0.8pt, mark=*, mark options={solid, darkgray}, forget plot]
  table[]{
  1	1
  1	2
  1	3
  1	4
  2	1
  2	2
  2	3
  2	4
  3	1
  3	2
  3	3
  3	4
  3	5
  3	6
  3	7
  3	8
  4	1
  4	2
  4	3
  4	4
  4	5
  4	6
  4	7
  4	8
  5	3
  5	4
  5	5
  5	6
  5	7
  5	8
  6	3
  6	4
  6	5
  6	6
  6	7
  6	8
  7	3
  7	4
  7	5
  7	6
  7	7
  7	8
  7	9
  7	10
  7	11
  7	12
  8	3
  8	4
  8	5
  8	6
  8	7
  8	8
  8	9
  8	10
  8	11
  8	12
  9	7
  9	8
  9	9
  9	10
  9	11
  9	12
  10	7
  10	8
  10	9
  10	10
  10	11
  10	12
  11	7
  11	8
  11	9
  11	10
  11	11
  11	12
  11	13
  11	14
  11	15
  11	16
  12	7
  12	8
  12	9
  12	10
  12	11
  12	12
  12	13
  12	14
  12	15
  12	16
  13	11
  13	12
  13	13
  13	14
  13	15
  13	16
  14	11
  14	12
  14	13
  14	14
  14	15
  14	16
  15	11
  15	12
  15	13
  15	14
  15	15
  15	16
  15	17
  15	18
  15	19
  15	20
  16	11
  16	12
  16	13
  16	14
  16	15
  16	16
  16	17
  16	18
  16	19
  16	20
  17	15
  17	16
  17	17
  17	18
  17	19
  17	20
  18	15
  18	16
  18	17
  18	18
  18	19
  18	20
  19	15
  19	16
  19	17
  19	18
  19	19
  19	20
  19	21
  19	22
  19	23
  19	24
  20	15
  20	16
  20	17
  20	18
  20	19
  20	20
  20	21
  20	22
  20	23
  20	24
  21	19
  21	20
  21	21
  21	22
  21	23
  21	24
  22	19
  22	20
  22	21
  22	22
  22	23
  22	24
  23	19
  23	20
  23	21
  23	22
  23	23
  23	24
  23	25
  23	26
  23	27
  23	28
  24	19
  24	20
  24	21
  24	22
  24	23
  24	24
  24	25
  24	26
  24	27
  24	28
  25	23
  25	24
  25	25
  25	26
  25	27
  25	28
  26	23
  26	24
  26	25
  26	26
  26	27
  26	28
  27	23
  27	24
  27	25
  27	26
  27	27
  27	28
  27	29
  27	30
  27	31
  27	32
  28	23
  28	24
  28	25
  28	26
  28	27
  28	28
  28	29
  28	30
  28	31
  28	32
  29	27
  29	28
  29	29
  29	30
  29	31
  29	32
  30	27
  30	28
  30	29
  30	30
  30	31
  30	32
  31	27
  31	28
  31	29
  31	30
  31	31
  31	32
  31	33
  31	34
  31	35
  31	36
  32	27
  32	28
  32	29
  32	30
  32	31
  32	32
  32	33
  32	34
  32	35
  32	36
  33	31
  33	32
  33	33
  33	34
  33	35
  33	36
  34	31
  34	32
  34	33
  34	34
  34	35
  34	36
  35	31
  35	32
  35	33
  35	34
  35	35
  35	36
  35	37
  35	38
  36	31
  36	32
  36	33
  36	34
  36	35
  36	36
  36	37
  36	38
  37	35
  37	36
  37	37
  37	38
  38	35
  38	36
  38	37
  38	38
};
\end{axis}
\end{tikzpicture}%

%% file: example.tex
% !TeX spellcheck = en_US
\section{Numerical Example and Evaluation}
\label{sec:example}

\tikzstyle{style1}=[black,dashed,line width=1.25pt]%,mark=*,mark options={solid,fill=black,draw=black,scale=0.6}]
\tikzstyle{style2}=[darkgray,dashed,line width=1.25pt] %,mark=*,mark options={solid,fill=black,draw=black,scale=0.6}]
\tikzstyle{style3}=[lightgray,dashed,line width=1.25pt] %,mark=*,mark options={solid,fill=darkgray,draw=darkgray,scale=0.6}]
\tikzstyle{style4}=[black,solid,line width=1.25pt]%,mark=*,mark options={solid,fill=black,draw=black,scale=0.6}]
\tikzstyle{style5}=[darkgray, solid, line width=1.25pt] %,mark=*,mark options={solid,fill=lightgray,draw=lightgray,scale=0.6}]
\tikzstyle{style6}=[lightgray, solid, line width=1.25pt] %,mark=*,mark options={solid,fill=lightgray,draw=lightgray,scale=0.6}]
\tikzstyle{style7}=[black, dash pattern={on 7pt off 2pt on 1pt off 3pt},line width=1.25pt]%,mark=*,mark options={solid,fill=black,draw=black,scale=0.6}]
\tikzstyle{style8}=[darkgray, dash pattern={on 7pt off 2pt on 1pt off 3pt},line width=1.25pt] %,mark=*,mark options={solid,fill=lightgray,draw=lightgray,scale=0.6}]
\tikzstyle{style9}=[lightgray, dash pattern={on 7pt off 2pt on 1pt off 3pt},line width=1.25pt] %,mark=*,mark options={solid,fill=lightgray,draw=lightgray,scale=0.6}]

In the following, a numerical example demonstrates the presented time-optimal control techniques. % in connection with a dual-mode realization.
The Van der Pol oscillator constitutes a second-order dynamic system with nonlinear damping and is commonly reported in the literature as benchmark system for control resp. system analysis methods.
Its dynamics are described by $\ddot{y}(t) - \big( 1 - y(t)^2 \big) \dot{y}(t) + y(t) = u(t)$ with $\fundef{y}{\mathbb{R}}{\mathbb{R}}$.
By defining the state vector $\xc{t} \mdef \big(\xcel{t}{1}, \xcel{t}{2}\big)^\transpose$, the nonlinear control-affine state space model according to~\eqref{eq:system_dynamics_cont} is given by:
\begin{equation}
\begin{split}
\xcdot{t}  &= \f\big(\xc{t}, u(t)\big) \\
&= \big(\xcel{t}{2},\ \big( 1 - \xcel{t}{1}^2 \big) \xcel{t}{2} - \xcel{t}{1} + u(t)\big)^\intercal. % TODO: vektor klammern?
\end{split}
\label{eq:vdp_system}
\end{equation}
Hereby, the unrestricted state and control sets are $\xset = \mathbb{R}^2$ and $\uset = \mathbb{R}$. 
For a given control reference $\urefscalar \in \uset$, the system exhibits a unique steady state at $\xeq = (\urefscalar, 0)^\transpose$.
Refer to~\cite{james1974_ima} for a detailed control synthesis.
%Note, for the simulations and experiments in this thesis, controllability and reachability % TODO definitionen
%are ensured by obtaining feasible solutions from the nonlinear program solvers satisfying the first-order optimality conditions. % TODO: vielleicht können wir das mit der Quelle doch eingrenzen auf unsere Anwendungsbereiche??!
%As a result, the limit cycle remains active and a potential steady state with  $|\urefscalar| < 1$ represents an unstable focus that makes stabilization more difficult.
In the following, the constraint sets are set to $\rxset =\xset$ and $\ruset = \{u \in \uset \mid |u| \leq 1\}$ respectively.

\subsection{Open-Loop Control}

This section investigates open-loop control in terms of a comparative analysis. 
Boundary values for the optimal control task are set to $\xs=(0,0)^\transpose$ and $\xf=(0.8,0)^\transpose$.
The solution to the initial value problem~\eqref{eq:system_dynamics_cont} with system~\eqref{eq:vdp_system} is approximated with forward Euler.

\begin{figure}[tb]
	\centering
	%\tikzsetnextfilename{grid_comparison_exact_fe_vdp}%
	\input{grid_comparison_exact_fe.tex}
	\caption{Comparison of the grids with forward Euler for the Van der Pol oscillator. IPOPT with explicit Hessian computation is shown at the top and SQP at the bottom.}
	\label{fig:benchmark:open_loop:grid_comparison_exact_fe:vdp}
\end{figure}
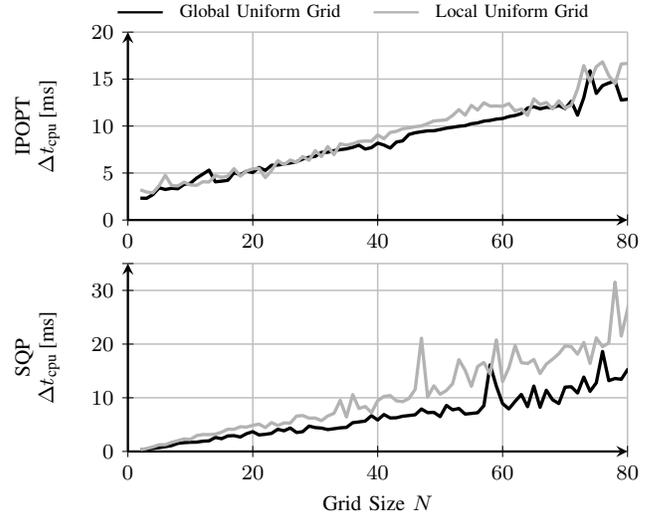

The first analysis compares the global and local uniform grid performances for two different solver configurations that both exploit sparsity using the hypergraph and internal sparse algebra.
In the first configuration, the nonlinear programs are solved by the sparse interior point solver IPOPT~\cite{waechter2006_matprog} and HSL-MA57 as internal linear solver~\cite{hsl}.
The Jacobian is computed via sparse finite differences and a step width of \num{1e-9}.
Note that we also compute the explicit Hessian of the Lagrangian with two consecutive Jacobians and step width \num{1e-2}.
The second configuration is based on a sequential quadratic programming (SQP) approach. 
The underlying quadratic program solver is the sparse general purpose solver %based on the alternating direction method of multipliers called
OSQP~\cite{osqp}.
Our SQP method squares objective terms $\vardt_k$ in~\eqref{eq:local_uniform_nlp} resp. $\vardt$ in \eqref{eq:global_uniform_nlp} and utilizes the Hessian of the objective rather than the Hessian of the Lagrangian.
This procedure ensures positive definiteness of the Hessians required for OSQP without changing the actual minimizer. Furthermore, it speeds up computation times since additional constraint evaluations are omitted.
The source code is available as part of our open-source C++ predictive control framework~\cite{corbo}.
\reffigc{fig:benchmark:open_loop:grid_comparison_exact_fe:vdp} shows the median computation times for both grid realizations and varying grid size $N$ (Ubuntu 16.04, Intel Core i7-4770 CPU at \SI{3.4}{GHz}, \SI{8}{GB} RAM, 20 repetitions).
The computation times are almost comparable. For SQP, the global uniform grid performs slightly better. 
It is noticeable that due to the hypergraph the computation time increases almost linearly with the grid size. 

%\OCPc{}~\eqref{eq:global_uniform_nlp} with system~\eqref{eq:vdp_system} is solved by the sparse interior point solver IPOPT~\cite{waechter2006_matprog} and HSL-MA57 as internal linear solver~\cite{hsl} in C++. 
%The implementation provides numerically computed sparse Jacobian and Hessian matrices according to the hypergraph formulation proposed in~\cite{roesmann2018_aim}. 
%Note that the hypergraph simplifies the determination of sparsity patterns during runtime and is particularly suitable for nonlinear programs with variable structures, e.g. as a result of grid adaptation. 
%The solution to the initial value problem~\eqref{eq:system_dynamics_cont} with system~\eqref{eq:vdp_system} is approximated with forward Euler.

% However, for small to mid size problems the computation times are comparable to each other and in both cases increase almost linearly in relation to the grid size. 

\setlength{\tabcolsep}{1pt}
\renewcommand{\arraystretch}{0.96}
\begin{table}[tb]
	\caption{Open-loop solving times in \si{ms} and integral errors for the Van der Pol oscillator with varying grid sizes $N$. The reference grid size which corresponds to $\vardt \approx \SI{0.1}{s}$ is indicated by $N^*$.}
	\label{tab:benchmark:tompc_l1stab}
	\notsotiny % TODO: really?
	\sisetup{round-mode=places,round-precision=2,table-format = 4.2}
	\begin{tabularx}{\columnwidth}{p{1.5cm} Y Y Y Y Y Y Y Y}
		\toprule
		& \multicolumn{2}{c}{$N = 5$} & \multicolumn{2}{c}{$N^* = 16$} & \multicolumn{2}{c}{$N = 25$} & \multicolumn{2}{c}{$N = 50$} \\
		\cmidrule(lr){2-3}\cmidrule(lr){4-5}\cmidrule(lr){6-7}\cmidrule(lr){8-9}
		& $\tcpumed$ & $\ierrtf$ & $\tcpumed$ & $\ierrtf$ & $\tcpumed$ & $\ierrtf$ & $\tcpumed$ & $\ierrtf$ \\
		\midrule
		TOMPC              				& \num{135.2699} & \num{0.055264} & \num{13.8778} & \num{0.055264} & \num{49.2511} & \num{0.055264} & \num{174.1594} & \num{0.055264} \\
		$\ell_1$-Norm ($\theta=1.01$) 	& ---			 & --- 		      & \num{4.0512}  & \num{0.04}     & \num{7.3767}  & \num{0.041777} & \num{9.8796}   & \num{0.041757} \\ % TODO just two places for N* since the last step is not perfectly zero...
		$\ell_1$-Norm ($\theta=1.1$)  	& ---  			 & ---	          & \num{4.4572}  & \num{0.04}     & \num{6.1364}  & \num{0.041752} & \num{12.8787}  & \num{0.041751} \\
		$\ell_1$-Norm ($\theta=1.5$)  	& --- 			 & ---	 	      & \num{8.8094}  & \num{0.04}     & \num{9.9065}  & \num{0.041752} & \num{30.1174}  & \num{0.041752} \\
		Local Grid 				& \num{3.1792}   & \num{0.070271} & \num{3.5198}  & \num{0.035659} & \num{4.4546}  & \num{0.019548} & \num{7.5824}   & $\approx 0$ \\ % $\approx$\num{0.0014797}  % TODO: last value is 0?! rounded
		\bottomrule
	\end{tabularx}
	% TODO: version mit konfidenzintervall in den anhang!
	% TODO: warum der fehler nicht gleich ist bei local uniform und l1 stab ist im bild zu erkennen. L1 stab beinhaltet schon die umschaltung auf 0! was ja eigentlich gut ist!!!!
	% -> achtung laufzeit geht runter bei l1 norm
\end{table}

The second analysis compares the proposed method with the state of the art approaches TOMPC and the $\ell_1$-norm approach as mentioned in \refsec{sec:introduction}.
The control task is as before and the local uniform grid is selected as the candidate for variable discretization.
TOMPC and the $\ell_1$-norm approach are configured with a fixed grid of resolution $\vardt = \SI{0.1}{s}$.
Recall that TOMPC adapts the grid size $N$ until the (quasi) minimum-time feasible solution is found which is for $N^*=16$ in this scenario.
Similarly, the $\ell_1$-norm approach requires at least a grid size of $N^*=16$ to return feasible solutions.
An advantage of the proposed variable discretization methods is that they may return a feasible solution even for $N < N^*$ while reducing accuracy.
To highlight this effect, we define an integral dynamics error w.r.t. the optimal solution $\xref$ obtained from a large grid resolution: 
$\ierrtf = \int_{0}^{\tf^*} \lVert x_\mtext{ref}(\tau) - \xsolfun\big(\tau, \xs, \uustar(\tau)\big)\rVert_2 \,\dtau.$
\reftab{tab:benchmark:tompc_l1stab} lists the medians of the computation times and dynamic errors for the IPOPT solver configuration.
Note that the $\ell_1$-norm approach requires to choose a design parameter $\theta$~\cite{verschueren2017_cdc}.
Choosing $\theta$ too large leads to fast growing values in the cost function which results in ill-conditions problems especially for large $N$ and hence larger computation times.
%Notice that for larger values of θ the computation times increase, although the optimization results respectively errors exˆ(tf) remain unchanged
The local uniform grid reveals the lowest computation times for all grid sizes in this scenario.

\subsection{Closed-Loop Control with Dual-Mode}

\begin{figure*}
	\centering
	\begin{subfigure}[b]{\columnwidth}
		\centering
		%\tikzsetnextfilename{ecp_closed_loop_states_legend}%
		\begin{tikzpicture}[font=\footnotesize]
		\begin{axis}[hide axis,	xmin=0,	xmax=1,	ymin=0,	ymax=1,	legend columns = 2,	legend pos = outer north, legend style={draw=none,fill=none,legend cell align=left, column sep =4pt, font=\scriptsize},
		combo legend1/.style={
			legend image code/.code={
				\draw [style1] (0cm,0cm) -- (0.6cm,0cm);
				\draw plot coordinates {(0.55cm,-3pt)} node {,};
				\draw [style2] (0.85cm,0cm) -- (1.55cm,0cm);
				\draw plot coordinates {(1.5cm,-3pt)} node {,};
				\draw [style3] (1.85cm,0cm) -- (2.45cm,0cm);
				%\draw [style3] plot coordinates {(4.5mm,0cm)};
			}
		},
		combo legend2/.style={
			legend image code/.code={
				\draw [style4] (0cm,0cm) -- (0.6cm,0cm);
				\draw plot coordinates {(0.55cm,-3pt)} node {,};
				\draw [style5] (0.85cm,0cm) -- (1.55cm,0cm);
				\draw plot coordinates {(1.5cm,-3pt)} node {,};
				\draw [style6] (1.85cm,0cm) -- (2.45cm,0cm);
			}
		},
		]
		\addlegendimage{combo legend2}; \addlegendentry{Time-Optimal};
		\addlegendimage{combo legend1}; \addlegendentry{Dual-Mode};
		\end{axis}
		\end{tikzpicture}
	\end{subfigure}	\\
	\begin{subfigure}{.5\textwidth}
		\centering
		%\tikzsetnextfilename{mpc_closed_loop_ecp_states}%
		%\input{images/dual_mode_nmin2/vdp_dualmode.tex}
		\input{vdp_dualmode.tex}
		\caption{}
		\label{fig:dual_mode_large}
	\end{subfigure}%
	\begin{subfigure}{.5\textwidth}
		\centering
		%\tikzsetnextfilename{mpc_closed_loop_ecp_states}%
		%\input{images/dual_mode_nmin2/vdp_dualmode_small.tex}
		\input{vdp_dualmode_small.tex}
		\caption{}
		\label{fig:dual_mode_small}
	\end{subfigure}%
	
	\caption{Closed-loop results for the Van der Pol oscillator. (a) Three individual solutions with $N=50$, $\Nmin=3$ and $\vardtmax=\SI{0.05}{s}$
		from different initial states (indicated by circles) are shown and the steady state $\xf = (0, 0)^\transpose$
		is marked by a cross. $P=\Pfullbig{\Nmin}{(\Nmin-1)\vardtmax}$ denotes the worst-case practical stability region with $\Nmin=3$. (b) Magnified views for the vicinity of $\xf$. For comparison, $P_2$ indicates the practical stability region for $\Nmin=2$.}
	\label{fig:dual_mode}
\end{figure*}
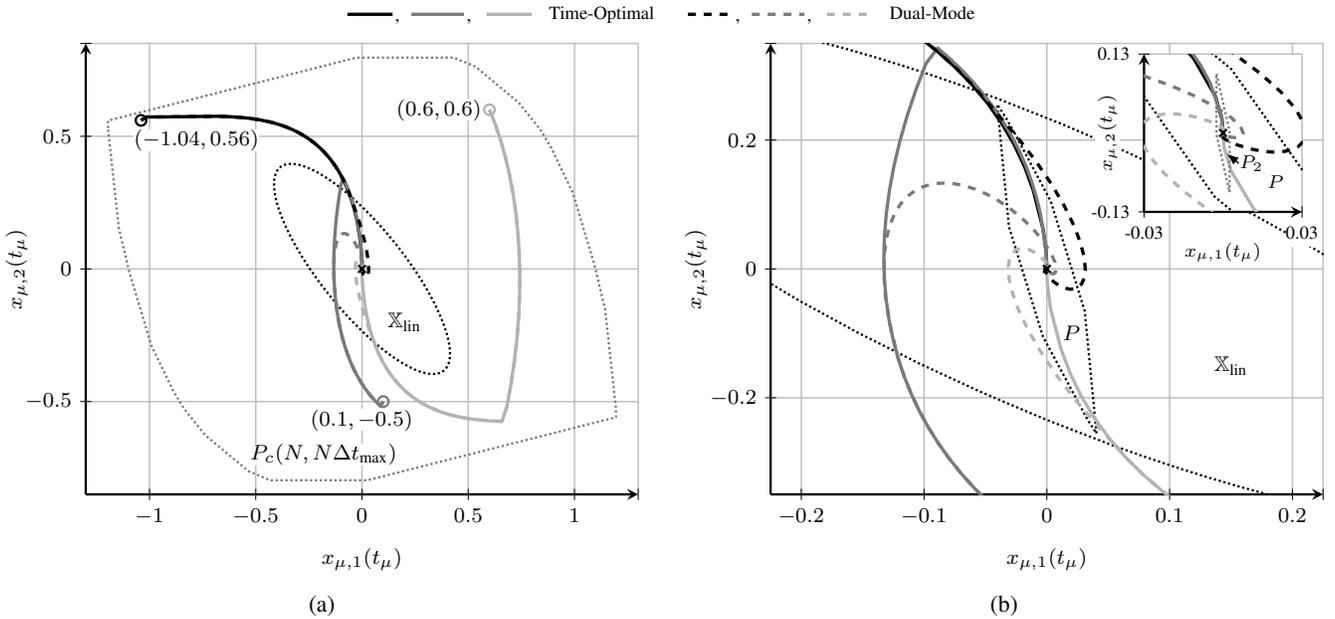%

% TODO: tfmin ist variable bei shrinking horizon!!!!!! , eigentlich ist dtmin fix!!!

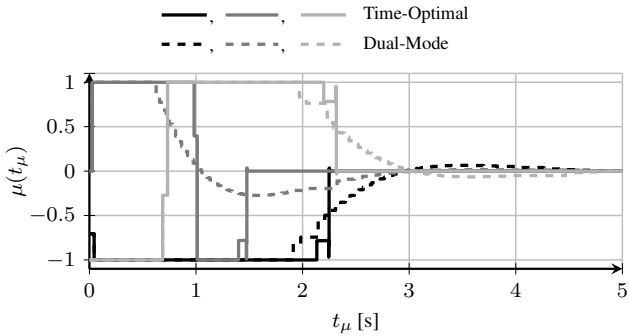
\begin{figure}[tb]
	\centering
	\begin{subfigure}[b]{\columnwidth}
		\centering
		%\tikzsetnextfilename{ecp_closed_loop_states_legend}%
		\begin{tikzpicture}[font=\footnotesize]
		\begin{axis}[hide axis,	xmin=0,	xmax=1,	ymin=0,	ymax=1,	legend columns = 1,	legend pos = outer north, legend style={draw=none,fill=none,legend cell align=left, column sep =4pt, font=\scriptsize},
		combo legend1/.style={
			legend image code/.code={
				\draw [style1] (0cm,0cm) -- (0.6cm,0cm);
				\draw plot coordinates {(0.55cm,-3pt)} node {,};
				\draw [style2] (0.85cm,0cm) -- (1.55cm,0cm);
				\draw plot coordinates {(1.5cm,-3pt)} node {,};
				\draw [style3] (1.85cm,0cm) -- (2.45cm,0cm);
				%\draw [style3] plot coordinates {(4.5mm,0cm)};
			}
		},
		combo legend2/.style={
			legend image code/.code={
				\draw [style4] (0cm,0cm) -- (0.6cm,0cm);
				\draw plot coordinates {(0.55cm,-3pt)} node {,};
				\draw [style5] (0.85cm,0cm) -- (1.55cm,0cm);
				\draw plot coordinates {(1.5cm,-3pt)} node {,};
				\draw [style6] (1.85cm,0cm) -- (2.45cm,0cm);
			}
		},
		]
		\addlegendimage{combo legend2}; \addlegendentry{Time-Optimal};
		\addlegendimage{combo legend1}; \addlegendentry{Dual-Mode};
		\end{axis}
		\end{tikzpicture}
	\end{subfigure}	\\
	\begin{subfigure}{\columnwidth}
		\centering
		\input{vdp_dualmode_controls.tex}
	\end{subfigure}%
	\caption{Evolution of the control inputs for the closed-loop realizations described in~\reffig{fig:dual_mode}.} % TODO: bis 5s simulieren
	\label{fig:dual_mode_controls}	
\end{figure}

This section investigates stabilizing closed-loop control with dual-mode.
The grid size is set to $N=50$ and $\Nmin=3$ serves as lower bound for the grid adaptation.
Setting $\vardtmax=\SI{0.05}{s}$ ensures the desired accuracy of the dynamics approximation and hence restricts the feasible state space for initial state $\xs$ to $\xs \in \Pfull{N}{N\vardtmax}$.
To account for a minimum grid resolution and closed-loop sampling times, $\vardtmin$ is set to $\vardtmin=\SI{1e-3}{s}$.
The control task consists of reaching steady state $\xf = (0, 0)$ from three different initial states $\xs\in\{ (-1.04, 0.56)^\transpose, (0.1, -0.5)^\transpose, (0.6,0.6)^\transpose \}$. % TODO: xf = 0 ?

For simplicity, the worst-case region of convergence $P=\Pfullbig{\Nmin}{(\Nmin-1)\vardtmax}$ is determined by sampling the control and time space using the reverse-time formulation of system~\eqref{eq:vdp_system}.
Note that $\vardtstar(\xs,N) \leq \vardtmax$ holds for all feasible initial states $\xs$.
Sampling is performed according to a 4-dimensional grid with steps of length~$0.1$ in the control and $\SI{0.01}{s}$ for the transition time.

%The linear controller for dual-mode operation is obtained by linearizing~\eqref{eq:vdp_system} at $\xf=(0,0)^\transpose$: % TODO: oder discrete time system?
Dual-mode operation is achieved by choosing a discrete-time linear quadratic regulator (LQR) as secondary feedback controller.
Hereby, we choose $\vardt_\mtext{LQR}=\vardtmin$ as sample time.
The design of the LQR requires a state error weighting matrix $R \in \mathbb{R}^{p \times p}$, a control error weighting matrix $Q \in \mathbb{R}^{q\times q}$ and a linear model $\x_{k+1} = A x_k + B u_k$ which follows from
linearizing the continuous-time model~\eqref{eq:vdp_system} at $\xf=(0,0)^\transpose$ and applying the zero-order hold method. 
The region of attraction $\rxsetlin$ is obtained by performing closed-loop simulations with the nonlinear system \eqref{eq:vdp_system}
and feedback $\mulawlin\big(\xmuc{\tmu}\big) = -K \xmuc{\tmu}$ according to a predefined grid of resolution $0.2$.
$\rxsetlin$ is then set to the largest inscribed ellipse.
Related parameters are listed below:
\begin{align*}
	A &= \bigl(\begin{smallmatrix} 1 & 0.001 \\ -0.001 & 1.001 \end{smallmatrix}\bigr), B = \bigl(\begin{smallmatrix} 0 \\ 0.001 \end{smallmatrix}\bigr), Q = \bigl(\begin{smallmatrix} 2 & 0 \\ 0 & 0.1 \end{smallmatrix}\bigr),
	R = (\begin{smallmatrix}0.1\end{smallmatrix}), \\
	K &\approx (\begin{smallmatrix} 3.5737 &   3.8711 \end{smallmatrix}), \rxsetlin = \{ x\mkern1.8mu{\in}\mkern1.8mu \xset \mid x^\transpose \bigl(\begin{smallmatrix} 16.65 & 14.03 \\ 14.03  & 18.19 \end{smallmatrix}\bigr) x \leq 1 \}.
	%\label{eq:system:ecpA:linss}
\end{align*} 
The linear system is controllable and hence stabilizable since $(A, AB)$ has full row rank.
Note that $Q$ and $R$ are chosen such that $\rxsetlin$ ensures $P \subseteq \rxsetlin$.

%\OCPc{}~\eqref{eq:global_uniform_nlp} with system~\eqref{eq:vdp_system} is solved by the sparse interior point solver IPOPT~\cite{waechter2006_matprog} and HSL-MA57 as internal linear solver~\cite{hsl} in C++. 
%The implementation provides numerically computed sparse Jacobian and Hessian matrices according to the hypergraph formulation proposed in~\cite{roesmann2018_aim}. 
%Note that the hypergraph simplifies the determination of sparsity patterns during runtime and is particularly suitable for nonlinear programs with variable structures, e.g. as a result of grid adaptation. 
%The solution to the initial value problem~\eqref{eq:system_dynamics_cont} with system~\eqref{eq:vdp_system} is approximated with forward Euler.

\reffigc{fig:dual_mode} shows the closed-loop simulation results for both the full time-optimal and dual-mode realizations. 
In addition, \reffig{fig:dual_mode_large} visualizes the feasibility region $\Pfull{N}{N\vardtmax}$. 
The enlarged views in \reffig{fig:dual_mode_small} also highlight the practical stability region for $\Nmin=3$
and for comparison also $\Nmin=2$ (marked by $P_2$), which is significantly smaller. Note that choosing $\Nmin=3$ is rather conservative in this example with $p=2$.
The three full time-optimal realizations are still able to stabilize the system even though recursive feasibility cannot be guaranteed.
In addition, the solution to~\eqref{eq:global_uniform_nlp} for states within the region of constant cost, i.e. for $\xmu \in \Pfull{\Nmin}{\Nmin\vardtmin}$
results in $\uref \mdef \uucstarbig{0,\xmu, {\Nmin}} = 0$ ensuring $f(\xmu,\uref)=0$. This does not hold for arbitrary systems and configurations
(refer also to the proofs of \refprop{shrinking_horizon_mpc:asymp_stability}). % and \reflem{stability_n1_kkt_conditions}). % TODO: besser formulieren?
The dual-mode controller switches to different trajectories as soon as $\rxsetlin$ is reached and ensures asymptotic stability.
\reffigc{fig:dual_mode_controls} depicts the evolution of control inputs associated with the closed-loop realizations.
Notice that each full time-optimal realization reveals a peak in control before switching to $\uref$. 
These peaks occur within the region $P$ as a result of the changed grid resolution and hence indicate potential recursive feasibility losses.
The dual-mode realization inherently leads to longer transition times, however, further tuning of $Q, R$ and $N$ affecting the sizes of $\rxsetlin$ and $P$ leads to more conservative respectively aggressive transitions. % TODO: mehr erklärung was N bewirkt?
Note that asymptotic stability also holds if the controller switches in $\rxsetlin=P$ (as $P \subseteq \rxsetlin$), assuming forward invariance of $P$, which significantly reduces transition times.

% TODO: initialization?
%An initialization of state and control trajectories is necessary for any derivative based nonlinear program solver. 
%In the very first invocation of \refalg{alg:uniform:shrinkinghorizon} the set of optimization parameters $\optimparamglobal{N}$ is empty.
%For the examples and applications in the scope of this thesis, a rather simple initialization strategy is utilized which can be applied to a wide range of small- and midscale optimal control problems. % TODO: müssen wir dafür eine analyse liefern?
%The state trajectory is approximated by linear interpolation between start state $\xmud{\tmud{n}}$ and final state $\xf$ with $N=\Ninit$ grid partitions, in particular
%\begin{equation}
%\xud{k} = \xmud{\tmud{n}} + \frac{k}{\Ninit}\big(\xf-\xmud{\tmud{n}}\big) \text{ for } k=0,1,\dotsc,\Ninit. % TODO: was ist mit multiple shooting, wo wir "s" anstatt "x" haben?
%\label{eq:uniform:gridadapt:init}
%\end{equation}
%The control sequence is initialized with zero controls $\uud{k}=\nullvec$ for $k=0,1,\dotsc,\Ninit$.
%In case of Hermite-Simpson collocation, initialization applies to $k=0,0.5,1,\dotsc,\Ninit$ respectively.
%The time resolution is initialized with some expected $\vardt > \vardtmin$.
%Obviously, this initialization strategy does not guarantee global convergence in the general case, since solving nonlinear programs numerically leads to a local minimizer with respect to the initialization $\optimparamglobal{N}$.
%% TODO hinweis auf praktische vorgehensweisen in der literatur: zunächst vereinfachtes problem lösen und damit das komplexe initialisieren.

\section{ECP Industrial Plant Emulator}
\label{sec:ecp}

\begin{figure}[tb]
	\centering
	%\vspace{-0.3cm}
	\includegraphics[width=8cm]{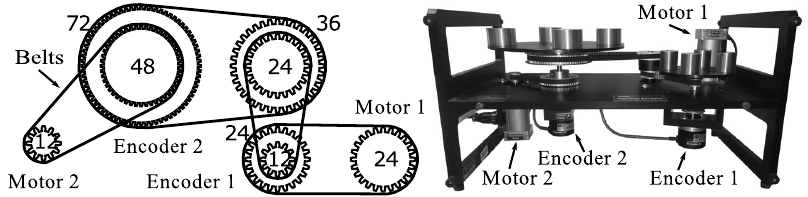}
	\vspace{-0.05cm}
	\caption{ECP Industrial Plant Emulator Model 220}
	\label{fig:ecp_model_picture}
	%\vspace{-0.3cm}
\end{figure}

This section investigates the closed-loop control on a real system as shown in Fig.~\ref{fig:ecp_model_picture}.
The \textit{ECP Industrial Plant Emulator Model 220} consists of two load plates actuated by motors which motion is coupled by transmission belts. 
Angular position and angular velocity are estimated from encoder signals with a DSP.

In the experimental setup, both motors generate torques to regulate the position and velocity of the plate of the secondary drive.
The system is described by the following nonlinear differential equation:
%\small
\begin{equation}
\ddot{x}(t) = -c_1 \dot{x}(t) - c_2 \tanh\big( c_3 \dot{x}(t) \big) + k_1 u_1(t) - k_2 u_2(t)
\end{equation}
%\normalsize
with $k_1= 34.51$, $k_2= 34.13$, $c_1 = 1.46$, $c_2 = 2.53$ and $c_3 = 5$.
Note, that $\tanh(\cdot)$ is chosen as a smooth approximation of the actual sign function.
%The state space model is constructed similar to Section~\ref{sec:vdp} with state vector $\mathbf{x}(t)=[x(t), \dot{x}(t)]^\intercal$.  
%Furthermore, the control inputs are limited to $|u(t)| \leq 0.5$ while the velocity is bounded to $|\dot{x}(t)| \leq 5$. 
The optimal control problem is constructed as before with state vector $\xc{t}=(x(t), \dot{x}(t))^\transpose$,  
control bounds $|u_1(t)| \leq 0.5$, $|u_2(t)| \leq 0.5$ and velocity bounds $|\dot{x}(t)| \leq 5$.
%Note, that the simplified dynamic equations exhibit a non-trivial model mismatch w.r.t. the true coupled drive dynamics.

The time optimal control task is specified with forward Euler integration, $\vardtref=0.05, \vardtmin = 0.001, \Nmin=2$ and $\xfset = \{\xf\}$ with $\xf = (\xfel{1}, \xfel{2})^\transpose$.
The $\ell_1$-norm approach serves as reference with $N=40$ to ensure feasibility for all transitions (TOMPC is not real-time capable in this scenario). 
The local uniform grid starts at $N=20$ and the grid is adapted by linear search as described in \refrem{rem:grid_adaptation}.
Note that $\xfel{1}$ is subject to change during runtime.
The dual-mode controller design is subject to the following parameters:
\begin{align*}
A &= \bigl(\begin{smallmatrix} 0 & 1 \\ 0 & 12.63\tanh(5\xfel{2})-14.1 \end{smallmatrix}\bigr), B = \bigl(\begin{smallmatrix} 0 & 0\\ 34.51 & -34.13 \end{smallmatrix}\bigr), \\
R &= (\begin{smallmatrix}1 & 0 \\ 0 & 1\end{smallmatrix}), Q = \bigl(\begin{smallmatrix} 1 & 0 \\ 0 & 0.1 \end{smallmatrix}\bigr), K \approx (\begin{smallmatrix} 0.711 &  0.131 \\ -0.703 & -0.130 \end{smallmatrix}), \\
\rxsetlin &= \{ x\mkern1.8mu{\in}\mkern1.8mu \xset \mid x^\transpose \bigl(\begin{smallmatrix} 2.2 & 0.38 \\ 0.38  & 0.11 \end{smallmatrix}\bigr) x \leq 1 \}.
%\label{eq:system:ecpA:linss}
\end{align*} 
Note that the linear approximation $(A,B)$ does not depend on the angular reference position $\xfel{1}$.
Therefore, $\rxsetlin$ is translated to $\xf = (\xfel{1}, 0)^\transpose$ whenever $\xfel{1}$ changes. 
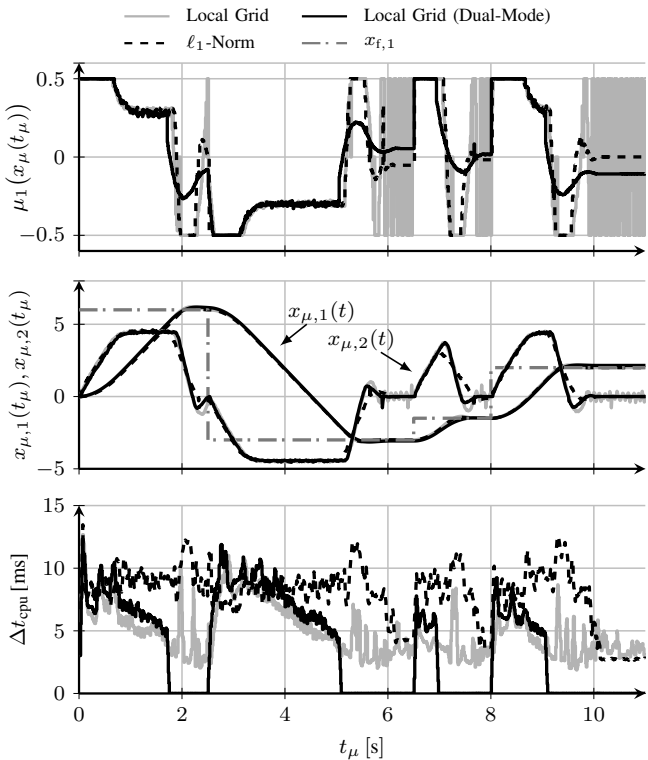
\begin{figure}[tb]
	\centering
	%\tikzsetnextfilename{benchmark_closed_loop_ecp_varying_ref}%
	\input{varying_reference.tex}
	\caption{Closed-loop control of the ECP Model 220 with varying final position $\ecpsf$. These are the local uniform grid, a dual-mode and hybrid cost realization and the $\ell_1$-norm approach. Note, $\mu_2(\cdot)$ is omitted as it is similar to $-\mu_1(\cdot)$.} 
	\label{fig:benchmark:closed_loop:ecp:varying_ref}
\end{figure}

\reffigc{fig:benchmark:closed_loop:ecp:varying_ref} shows the control, state and computation time profiles for the different realizations and varying reference positions $\xfel{1}$.
 The closed-loop performance between the local uniform grid (without dual-mode) and the $\ell_1$-norm approach is very similar which is reasonable because both claim to be time-optimal.
 According to the analysis in \refsec{sec:stability_analysis}, however, the controller does not guarantee stabilization, even if it is achieved here by chattering the control inputs.
 In contrast, the stabilizing quasi-time-optimal dual-mode realization performs quite similar at the beginning of each transition, but then lead to a smooth stabilization at $\xf$.
 Note that the drop in computation time (bottom plot) indicates when the LQR is active.
In this scenario, the computation times are comparable, but they decrease for the local uniform grid due to grid adaptation.

%% file: grid_comparison_exact_fe.tex
\setlength{\figurewidth}{0.75\columnwidth}
\setlength{\figureheight}{2.5cm}

\begin{tikzpicture}[font=\footnotesize] % TODO font size?!

\begin{axis}[%
width=\figurewidth,
height=\figureheight,
at={(1.011in,0.642in)},
name=plotipopt,
scale only axis,
xmin=0,
xmax=80,
ymin=0,
ymax=20,
ylabel={IPOPT\\$\tcpumed\,\mtext{[ms]}$},
y label style={at={(-0.12,0.5)},align=center},
axis background/.style={fill=white},
axis x line*=bottom,
axis y line*=left,
xmajorgrids,
ymajorgrids,
legend columns = 2,
legend pos = outer north,
legend style={draw=none,fill=none,legend cell align=left, column sep =4pt, font=\scriptsize,at={(\figurewidth/2-0.3cm,\figureheight+0cm)}},
]
\addplot [style4]
table[]{grid_comparison_exact_fe-1.tsv};
\addlegendentry{Global Uniform Grid}

\addplot [style6]
table[]{grid_comparison_exact_fe-2.tsv};
\addlegendentry{Local Uniform Grid}

%\addplot [style5]
%table[]{images/grid_comparison_vdp/tikz_data/grid_comparison_exact_fe-3.tsv};
%\addlegendentry{IPOPTe Quasiuniform FE}

%\addplot [style4]
%table[]{images/grid_comparison_vdp/tikz_data/grid_comparison_exact_fe-4.tsv};
%\addlegendentry{IPOPTe Timescaling FE}

\end{axis}

% SQP

\begin{axis}[%
width=\figurewidth,
height=\figureheight,
at=(plotipopt.below south west),
anchor=above north west,
scale only axis,
xmin=0,
xmax=80,
xlabel={Grid Size $N$},
ymin=0,
ymax=35,
extra y ticks={35},
extra y tick labels={},
ylabel={SQP\\$\tcpumed\,\mtext{[ms]}$},
y label style={at={(-0.12,0.5)},align=center},
axis background/.style={fill=white},
axis x line*=bottom,
axis y line*=left,
xmajorgrids,
ymajorgrids,
legend columns = 3,
legend pos = outer north,
legend style={draw=none,fill=none,legend cell align=left, column sep =4pt, font=\scriptsize,at={(\figurewidth/2-0.3cm,\figureheight+0cm)}},
]
\addplot [style4]
table[]{grid_comparison_sqp_fe-1.tsv};
%\addlegendentry{SQP Uniform G ED}

\addplot [style6]
table[]{grid_comparison_sqp_fe-2.tsv};
%\addlegendentry{SQP Uniform L ED}

%\addplot [style5]
%table[]{images/grid_comparison_vdp/tikz_data/grid_comparison_sqp_fe-3.tsv};
%\addlegendentry{SQP Quasiuniform ED}

%\addplot [style4]
%table[]{images/grid_comparison_vdp/tikz_data/grid_comparison_sqp_fe-4.tsv};
%\addlegendentry{SQP Timescaling ED}

\end{axis}

\end{tikzpicture}%

%% file: vdp_dualmode.tex
\definecolor{mycolor1}{rgb}{0.00000,0.44700,0.74100}%
\definecolor{mycolor2}{rgb}{0.85000,0.32500,0.09800}%
\definecolor{mycolor3}{rgb}{0.92900,0.69400,0.12500}%
\definecolor{mycolor4}{rgb}{0.49400,0.18400,0.55600}%
\definecolor{mycolor5}{rgb}{0.46600,0.67400,0.18800}%
\definecolor{mycolor6}{rgb}{0.30100,0.74500,0.93300}%
\begin{tikzpicture}[font=\footnotesize]

\begin{axis}[%
width=0.81\columnwidth,
height=6cm,
at={(2.6in,1.103in)},
scale only axis,
xmin=-1.3, % 1.3
xmax=1.3,
xlabel=$\xmucel{\tmu}{1}$,
extra x ticks={1.3},
extra x tick labels={},
x label style={at={(0.5,-0.1)}},
ymin=-0.85, %0.85
ymax=0.85,
ylabel=$\xmucel{\tmu}{2}$,
y label style={at={(-0.09,0.5)}},
extra y ticks={0.85},
extra y tick labels={},
axis background/.style={fill=white},
axis x line*=bottom,
axis y line*=left,
xmajorgrids,
ymajorgrids
]
\addplot [color=darkgray, densely dotted, forget plot] table[]{vdp_dualmode-1.tsv};
%\draw[draw=black] (axis cs:-1.5,-1) rectangle (axis cs:1.5,1);
\addplot [color=black, densely dotted, forget plot] table[]{vdp_dualmode-2.tsv};
%\addplot [color=black, forget plot] table[]{vdp_dualmode-3.tsv};

\addplot [style4, forget plot] table[]{vdp_dualmode-4.tsv}; % conf1
\addplot [style5, forget plot] table[]{vdp_dualmode-5.tsv}; % conf2
\addplot [style6, forget plot] table[]{vdp_dualmode-6.tsv}; % conf3
\addplot [style1, forget plot] table[]{vdp_dualmode-7.tsv}; % conf1 dm
\addplot [style2, forget plot] table[]{vdp_dualmode-8.tsv}; % conf2 dm
\addplot [style3, forget plot] table[]{vdp_dualmode-9.tsv}; % conf3 dm

% equilibrium
\addplot [color=black, draw=none, mark=x, mark options={solid, black}, forget plot] coordinates {(0,0)};

% start positions
\coordinate (xconf1) at (-1.04,0.56);
\coordinate (xconf2) at (0.1,-0.5);
\coordinate (xconf3) at (0.6,0.6);
\addplot [color=black, draw=none, mark=o, mark options={solid, black}, forget plot] coordinates {(-1.04,0.56)};
\addplot [color=darkgray, draw=none, mark=o, mark options={solid, darkgray}, forget plot] coordinates {(0.1,-0.5)};
\addplot [color=lightgray, draw=none, mark=o, mark options={solid, lightgray}, forget plot] coordinates {(0.6,0.6)};

\node[below right = 0.12cm and -0.1cm, fill=white, inner sep=0pt] at (xconf1) {$(-1.04,0.56)$};
\node[below left=0.1cm and -0.4cm, fill=white, inner sep=0pt] at (xconf2) {$(0.1,-0.5)$};
\node[left=0.1cm, fill=white, inner sep=0pt] at (xconf3) {$(0.6,0.6)$};

% draw region names
\node [fill=white, inner sep=0pt] at (-0.18, -0.7) {$\Pfull{N}{N\vardtmax}$};
%\draw[thin, latex-] (0.06, -0.15) -- node[pos=1, right, inner sep=0pt] {$\rxsetlin$} (0.25, -0.2);
\node[] at (0.2, -0.2) {$\rxsetlin$};
\end{axis}
\end{tikzpicture}%

%% file: vdp_dualmode_small.tex
\definecolor{mycolor1}{rgb}{0.00000,0.44700,0.74100}%
\definecolor{mycolor2}{rgb}{0.85000,0.32500,0.09800}%
\definecolor{mycolor3}{rgb}{0.92900,0.69400,0.12500}%
\definecolor{mycolor4}{rgb}{0.49400,0.18400,0.55600}%
\definecolor{mycolor5}{rgb}{0.46600,0.67400,0.18800}%
\definecolor{mycolor6}{rgb}{0.30100,0.74500,0.93300}%
\begin{tikzpicture}[font=\footnotesize]

\begin{axis}[%
width=0.81\columnwidth,
height=6cm,
at={(2.6in,1.103in)},
name=plot1,
scale only axis,
xmin=-0.225,
xmax=0.225,
xlabel=$\xmucel{\tmu}{1}$,
extra x ticks={0.225},
extra x tick labels={},
x tick label style={
	/pgf/number format/fixed,
	/pgf/number format/precision=5
},
scaled y ticks=false, % no decimal exponent
x label style={at={(0.5,-0.1)}},
ymin=-0.35,
ymax=0.35,
ylabel=$\xmucel{\tmu}{2}$,
y label style={at={(-0.1,0.5)}},
extra y ticks={0.35},
extra y tick labels={},
axis background/.style={fill=white},
axis x line*=bottom,
axis y line*=left,
xmajorgrids,
ymajorgrids
]
%\addplot [color=green, forget plot] table[]{images/dual_mode/tikz_data/vdp_dualmode-1.tsv};
\addplot [color=black, densely dotted, forget plot] table[]{vdp_dualmode-2.tsv};
\addplot [color=black, densely dotted, forget plot] table[]{vdp_dualmode-3.tsv};

\addplot [style4, forget plot] table[]{vdp_dualmode-4.tsv}; % conf1
\addplot [style5, forget plot] table[]{vdp_dualmode-5.tsv}; % conf2
\addplot [style6, forget plot] table[]{vdp_dualmode-6.tsv}; % conf3
\addplot [style1, forget plot] table[]{vdp_dualmode-7.tsv}; % conf1 dm
\addplot [style2, forget plot] table[]{vdp_dualmode-8.tsv}; % conf2 dm
\addplot [style3, forget plot] table[]{vdp_dualmode-9.tsv}; % conf3 dm

%\draw[draw=black] (axis cs:-1.5,-1) rectangle (axis cs:1.5,1);

% equilibrium
\addplot [color=black, draw=none, mark=x, mark options={solid, black}, forget plot] coordinates {(0,0)};

\node [] at (0.02, -0.1) {$P$};
\node [] at (0.15, -0.15) {$\rxsetlin$};

\end{axis}

% small preview
\begin{axis}[%
width=2.1cm,
height=2.1cm,
at=(plot1.north east),
anchor=above north east,
xshift=-0.28cm,
yshift=0.05cm,
scale only axis,
font=\scriptsize,
xmin=-0.03,
xmax=0.03,
xlabel={$\xmucel{\tmu}{1}$},
x label style={at={(0.5,-0.14)}},
%ticklabel style={fill=white},
xtick = {-0.03, 0.03}, % show box if majorgrids are on
xticklabels={-0.03, 0.03}, % hide x tick labels
restrict x to domain = -0.1:0.1,
x tick label style={
	/pgf/number format/fixed,
	/pgf/number format/precision=5,
	fill=white, inner xsep=0pt
},
scaled x ticks=false, % no decimal exponent
ymin=-0.13,
ymax=0.13,
ytick = {-0.13, 0.13}, % show box if majorgrids are on
yticklabels={-0.13,0.13}, % hide x tick labels
y tick label style={
	/pgf/number format/fixed,
	/pgf/number format/precision=5,
},
scaled y ticks=false, % no decimal exponent
ylabel={$\xmucel{\tmu}{2}$},
y label style={fill=white, inner sep=1pt, at={(-0.15,0.5)}},
%ticks=none, % hide ticks
%restrict y to domain = -1.1:-0.5,
axis background/.style={fill=white},
xmajorgrids,
ymajorgrids,
]
%\addplot [color=green, forget plot] table[]{images/dual_mode/tikz_data/vdp_dualmode-1.tsv};
\addplot [color=black, densely dotted, forget plot] table[]{vdp_dualmode-2.tsv};
\addplot [color=black, densely dotted, forget plot] table[]{vdp_dualmode-3.tsv};

\addplot [style4, forget plot] table[]{vdp_dualmode-4.tsv}; % conf1
\addplot [style5, forget plot] table[]{vdp_dualmode-5.tsv}; % conf2
\addplot [style6, forget plot] table[]{vdp_dualmode-6.tsv}; % conf3
\addplot [style1, forget plot] table[]{vdp_dualmode-7.tsv}; % conf1 dm
\addplot [style2, forget plot] table[]{vdp_dualmode-8.tsv}; % conf2 dm
\addplot [style3, forget plot] table[]{vdp_dualmode-9.tsv}; % conf3 dm

% equilibrium
\addplot [color=black, draw=none, mark=x, mark options={solid, black}, forget plot] coordinates {(0,0)};

\coordinate (pintl) at (-0.2, 0.5);
\coordinate (pinor) at (-0.2, -0.5);
\coordinate (pinbr) at (0.2, -0.5);

% Nmin=3
% draw practical stability region
%\addplot[forget plot, densely dotted, color=black] coordinates {(0.0549, -2.9271) (0.0549,-0.1814) (-0.0549,	2.9271)  (-0.0549,0.1814) (0.0549, -2.9271)};
%\draw[thin, latex-] (0.0014, -0.034) -- node[pos=1, right, inner sep=0pt] {$P$} (0.006, -0.05);
\node[] at (0.02, -0.08) {$P$};

% Nmin=2
% draw practical stability region
\addplot[forget plot, densely dotted, color=darkgray] coordinates {(0.0025, -0.0975) (0.0025,0.0025) (-0.0025,0.09754) (-0.0025,	-0.0025) (0.0025,-0.09754) };

\draw[thin, latex-] (0.0014, -0.034) -- node[pos=1, right, inner sep=0pt] {$P_2$} (0.006, -0.05);
\end{axis}
\end{tikzpicture}%

%% file: vdp_dualmode_controls.tex
\definecolor{mycolor1}{rgb}{0.00000,0.44700,0.74100}%
\definecolor{mycolor2}{rgb}{0.85000,0.32500,0.09800}%
\definecolor{mycolor3}{rgb}{0.92900,0.69400,0.12500}%
\definecolor{mycolor4}{rgb}{0.49400,0.18400,0.55600}%
\definecolor{mycolor5}{rgb}{0.46600,0.67400,0.18800}%
\definecolor{mycolor6}{rgb}{0.30100,0.74500,0.93300}%
\begin{tikzpicture}[font=\footnotesize]

\begin{axis}[%
width=0.8\columnwidth,
height=2.6cm,
at={(0.758in,0.481in)},
scale only axis,
xmin=0,
xmax=5,
%extra x ticks={4.2},
%extra x tick labels={},
%x label style={at={(0.5,0)}},
xlabel=$\tmu\,\mtext{[s]}$,
ymin=-1.1,
ymax=1.1,
extra y ticks={1.1},
extra y tick labels={},
y label style={at={(-0.1,0.5)}},
ylabel=$\mulaw(\tmu)$, % TODO or mu_Nd?
axis background/.style={fill=white},
xmajorgrids,
ymajorgrids
]
\addplot [style4, const plot, forget plot] table[]{vdp_dualmode_controls-1.tsv};
\addplot [style1, const plot, forget plot] table[]{vdp_dualmode_controls-2.tsv};
\addplot [style5, const plot, forget plot] table[]{vdp_dualmode_controls-3.tsv};
\addplot [style2, const plot, forget plot] table[]{vdp_dualmode_controls-4.tsv};
\addplot [style6, const plot, forget plot] table[]{vdp_dualmode_controls-5.tsv};
\addplot [style3, const plot, forget plot] table[]{vdp_dualmode_controls-6.tsv};
\end{axis}
\end{tikzpicture}%

%% file: varying_reference.tex
\setlength{\figurewidth}{0.85\columnwidth}
\setlength{\figureheight}{2.5cm}

\begin{tikzpicture}[font=\footnotesize] % TODO font size?!

\begin{axis}[%
width=\figurewidth,
height=\figureheight,
name=plotu1,
at={(1.011in,4.856in)},
scale only axis,
xmin=0,
xmax=11,
xticklabels={,,}, % hide x tick labels
ymin=-0.6,
ymax=0.6,
extra y ticks={0.6},
extra y tick labels={},
ylabel={$\mulaw_1\big(\xmuc{\tmu}\big)$},
y label style={at={(-0.07,0.5)}},
axis background/.style={fill=white},
axis x line*=bottom,
axis y line*=left,
xmajorgrids,
ymajorgrids,
legend columns = 2,
legend pos = outer north,
legend style={draw=none,fill=none,legend cell align=left, column sep =4pt, font=\scriptsize,at={(\figurewidth/2-0.3cm,\figureheight+0cm)}},
]
\addplot [const plot, style6] table[]{varying_reference-1.tsv}; \addlegendentry{Local Grid}
%\addplot [const plot, color=mycolor2] table[]{content/benchmark/varying_reference-2.tsv}; \addlegendentry{Global Uniform Grid}
%\addplot [const plot, color=mycolor3] table[]{content/benchmark/varying_reference-3.tsv}; \addlegendentry{Global Uniform Grid (Dual-Mode)}
\addplot [const plot, style4] table[]{varying_reference-4.tsv}; \addlegendentry{Local Grid (Dual-Mode)}
%\addplot [const plot, style2] table[]{varying_reference-5.tsv}; \addlegendentry{Hybrid Cost}
\addplot [const plot, style1] table[]{varying_reference-6.tsv}; \addlegendentry{$\ell_1$-Norm}

\addlegendimage{style8}; \addlegendentry{$\ecpsf$}; % TODO: only first component
\end{axis}

%\begin{axis}[%
%width=\figurewidth,
%height=\figureheight,
%name=plotu2,
%at=(plotu1.below south west),
%anchor=above north west,
%scale only axis,
%xmin=0,
%xmax=11,
%xticklabels={,,}, % hide x tick labels
%ymin=-0.6,
%ymax=0.6,
%ylabel={$\mulaw_2\big(\xmuc{\tmu}\big)$},
%y label style={at={(-0.07,0.5)}},
%axis background/.style={fill=white},
%axis x line*=bottom,
%axis y line*=left,
%xmajorgrids,
%ymajorgrids,
%]
%\addplot [const plot, style3, forget plot] table[]{content/benchmark/varying_reference-7.tsv}; % Local
%%\addplot [const plot, color=mycolor2, forget plot] table[]{content/benchmark/varying_reference-8.tsv}; % Global
%%\addplot [const plot, color=mycolor3, forget plot] table[]{content/benchmark/varying_reference-9.tsv}; % Dual-Mode Global
%\addplot [const plot, style1, forget plot] table[]{content/benchmark/varying_reference-10.tsv}; % Dual-Mode Local
%\addplot [const plot, style2, forget plot] table[]{content/benchmark/varying_reference-11.tsv}; % Hybrid
%\addplot [const plot, style4, forget plot] table[]{content/benchmark/varying_reference-12.tsv}; % l1 norm
%\end{axis}

\begin{axis}[%
width=\figurewidth,
height=\figureheight,
name=plotx1,
at=(plotu1.below south west),
anchor=above north west,
scale only axis,
xmin=0,
xmax=11,
xticklabels={,,}, % hide x tick labels
ymin=-5,
ymax=8,
extra y ticks={8},
extra y tick labels={},
ylabel={$\xmucel{\tmu}{1}, \xmucel{\tmu}{2}$},
y label style={at={(-0.07,0.5)}},
axis background/.style={fill=white},
axis x line*=bottom,
axis y line*=left,
xmajorgrids,
ymajorgrids
]
\addplot [style6, forget plot] table[]{varying_reference-13.tsv} % Local
node[pos=0.4,pin={[shift={(0.0cm,0.0cm)},inner sep=0pt,pin distance=0.3cm,color=black, pin edge={latex-, black, semithick}]60:$\xmucel{t}{1}$}, color=black]{};

%\addplot [color=mycolor2, forget plot] table[]{content/benchmark/varying_reference-14.tsv}; % Global
%\addplot [color=mycolor3, forget plot] table[]{content/benchmark/varying_reference-15.tsv}; % Dual Mode Global
\addplot [style4, forget plot] table[]{varying_reference-16.tsv}; % Dual Mode Local
%\addplot [style2, forget plot] table[]{varying_reference-17.tsv}; % Hybrid
\addplot [style1, forget plot] table[]{varying_reference-18.tsv}; % l1 norm
\addplot [const plot, style8, forget plot] table[]{varying_reference-19.tsv}; % reference

%% x2:

\addplot [style6, forget plot] table[]{varying_reference-20.tsv} % Local
node[pos=0.50,pin={[shift={(0.0cm,0.0cm)},inner sep=0pt,pin distance=0.3cm,color=black, pin edge={latex-, black, semithick}]135:$\xmucel{t}{2}$}, color=black]{};

%\addplot [color=mycolor2, forget plot] table[]{content/benchmark/varying_reference-21.tsv}; % Global
%\addplot [color=mycolor3, forget plot] table[]{content/benchmark/varying_reference-22.tsv}; % Dual Mode Global
\addplot [style1, forget plot] table[]{varying_reference-23.tsv}; % Dual Mode Local
%\addplot [style2, forget plot] table[]{varying_reference-24.tsv}; % Hybrid
\addplot [style4, forget plot] table[]{varying_reference-25.tsv}; % l1 norm

\end{axis}

%\begin{axis}[%
%width=\figurewidth,
%height=\figureheight,
%name=plotx2,
%at=(plotx1.below south west),
%anchor=above north west,
%scale only axis,
%xmin=0,
%xmax=11,
%xticklabels={,,}, % hide x tick labels
%ymin=-6,
%ymax=6,
%ylabel={$\xmucel{\tmu}{2}$},
%y label style={at={(-0.07,0.5)}},
%axis background/.style={fill=white},
%axis x line*=bottom,
%axis y line*=left,
%xmajorgrids,
%ymajorgrids
%]
%\addplot [style3, forget plot] table[]{content/benchmark/varying_reference-20.tsv}; % Local
%%\addplot [color=mycolor2, forget plot] table[]{content/benchmark/varying_reference-21.tsv}; % Global
%%\addplot [color=mycolor3, forget plot] table[]{content/benchmark/varying_reference-22.tsv}; % Dual Mode Global
%\addplot [style1, forget plot] table[]{content/benchmark/varying_reference-23.tsv}; % Dual Mode Local
%\addplot [style2, forget plot] table[]{content/benchmark/varying_reference-24.tsv}; % Hybrid
%\addplot [style4, forget plot] table[]{content/benchmark/varying_reference-25.tsv}; % l1 norm
%\end{axis}

\begin{axis}[%
width=\figurewidth,
height=\figureheight,
name=plotcpu,
at=(plotx1.below south west),
anchor=above north west,
scale only axis,
xmin=0,
xmax=11,
%xticklabels={,,}, % hide x tick labels
xlabel={$\tmu\,[\mtext{s}]$},
ymin=0,
ymax=15,
ylabel={$\tcpu\,\mtext{[ms]}$},
y label style={at={(-0.07,0.5)}},
axis background/.style={fill=white},
axis x line*=bottom,
axis y line*=left,
xmajorgrids,
ymajorgrids
]
\addplot [const plot, style6,  forget plot] table[]{varying_reference-26.tsv}; % Local
%\addplot [const plot, color=mycolor2, forget plot] table[]{content/benchmark/varying_reference-27.tsv}; % Global
%\addplot [const plot, color=mycolor3, forget plot] table[]{content/benchmark/varying_reference-28.tsv}; % Dual Mode Global
\addplot [const plot, style4,  forget plot] table[]{varying_reference-29.tsv}; % Dual Mode Local
%\addplot [const plot, style2,  forget plot] table[]{varying_reference-30.tsv}; % Hybrid
\addplot [style1,  forget plot] table[]{varying_reference-31.tsv}; % l1 norm
\end{axis}

%\begin{axis}[%
%width=\figurewidth,
%height=\figureheight,
%name=plotn,
%at=(plotcpu.below south west),
%anchor=above north west,
%scale only axis,
%xmin=0,
%xmax=11,
%xlabel={$\tmu\,[\mtext{s}]$},
%ymin=0,
%ymax=50,
%ylabel={$N$},
%y label style={at={(-0.07,0.5)}},
%axis background/.style={fill=white},
%axis x line*=bottom,
%axis y line*=left,
%xmajorgrids,
%ymajorgrids,
%]
%\addplot [const plot, style3] table[]{content/benchmark/varying_reference-32.tsv}; % Local
%%\addplot [const plot, color=mycolor2] table[]{content/benchmark/varying_reference-33.tsv}; % Global
%%\addplot [const plot, color=mycolor3] table[]{content/benchmark/varying_reference-34.tsv}; % Dual Mode Global
%\addplot [const plot, style1] table[]{content/benchmark/varying_reference-35.tsv}; % Dual Mode Local
%\addplot [const plot, style2] table[]{content/benchmark/varying_reference-36.tsv}; % Hybrid
%\addplot [const plot, style4] table[]{content/benchmark/varying_reference-37.tsv}; % l1 norm
%
%\end{axis}
\end{tikzpicture}%

%% file: summary.tex
% !TeX spellcheck = en_US
\section{Conclusion}
\label{sec:summary}

Common time-optimal control formulations with state feedback, for example based on time transformation, do not take the control parameterization typically arising in \MPC{} into account
such that recursive feasibility and stabilization can no longer be guaranteed.
To this end, this paper proposes a suitable formulation based on variable discretization and grid adaptation for which closed-loop convergence results are derived.
Even though these results mainly deal with the nominal case, however, they include design parameters, i.e. lower bounds on the temporal resolution and the grid size,
to inherently deal with numerical ill-conditioning and a potential loss of viability.
In addition, a dual-mode control scheme results in a smooth quasi-time-optimal stabilization. % while restoring asymptotic stability. 
From an implementation point of view, two proposed nonlinear program formulations result in the same optimal solution but reveal different sparsity patterns.
Under more restricted conditions and a class of numerical integration schemes, these formulations also ensure true time-optimality and asymptotic stability without dual-mode. % TODO: so lassen?
A numerical example demonstrates the design of the proposed schemes and provides a comparative analysis with state of the art approaches.  %compares the dual-mode realization with the time-optimal one.
Experimental results with a real system highlight the practical feasibility of the proposed method.

Future work investigates moving horizon time-optimal control schemes with terminal conditions that do not necessarily require a dedicated dual-mode design to asymptotically stabilize the system.
% TODO: LQR durch metaoptimierung zur einhaltung von P bringen